\documentclass[aps,prd,showpacs,twocolumn,superscriptaddress,nofootinbib,preprintnumbers]{revtex4-1} 

\makeatletter
\def\p@subsection{}
\makeatother

\usepackage{color,graphicx,amsmath,amssymb,lipsum}
\usepackage[colorlinks=true,citecolor=blue,linkcolor=blue,urlcolor=blue, backref=false,pdfborder={0 0 0}]{hyperref}

\usepackage[normalem]{ulem}
\usepackage[utf8]{inputenc} 
\usepackage{mathtools}
\usepackage{tabularx} % or tabular
\usepackage{tabu}
\usepackage{multirow}
\usepackage[table]{xcolor}
\usepackage{colortbl}
\usepackage{caption}
\usepackage{tikz}
\usepackage{bbm}
\usetikzlibrary{snakes}
\usetikzlibrary{decorations}
\usetikzlibrary{trees}
\usetikzlibrary{decorations.pathmorphing}
\usetikzlibrary{decorations.markings}
\usetikzlibrary{external}
\usetikzlibrary{intersections}
\usetikzlibrary{shapes,arrows}
\usetikzlibrary{arrows.meta}
\usetikzlibrary{calc}
\usetikzlibrary{shapes.misc}
\usetikzlibrary{decorations.text}
\usetikzlibrary{backgrounds}
\usetikzlibrary{fadings}
\usetikzlibrary{tikzmark,calc,arrows,shapes,decorations.pathreplacing}
\tikzset{
        cross/.style={cross out, draw=black, minimum size=2*(#1-\pgflinewidth), inner sep=0pt, outer sep=0pt},
	branchCut/.style={postaction={decorate},
		snake=zigzag,
		decoration = {snake=zigzag,segment length = 2mm, amplitude = 2mm}	
    }}

\usepackage{float}

\usepackage{bm}

\usepackage{amsmath}
\usepackage{tikz}
\usepackage{tikz-feynman}
\tikzfeynmanset{compat=1.1.0}

\include{figures}

\usepackage{ulem}
\usepackage{diagbox}

\usepackage{mathrsfs}

\def\fft#1#2{{\frac{#1}{#2}}}

\newcommand{\be}{\begin{equation}}
\newcommand{\ee}{\end{equation}}
\newcommand{\beqa}{\begin{eqnarray}}
\newcommand{\eeqa}{\end{eqnarray}}

\newcommand{\bseq}{\begin{subequations}}
\newcommand{\eseq}{\end{subequations}}

\definecolor{cornellRed}{HTML}{B31B1B}
\definecolor{cornellBlue}{HTML}{0068AC}
\definecolor{cornellGreen}{HTML}{6EB43F}
\definecolor{purple}{HTML}{66023C}

\def\gsim{\raise0.3ex\hbox{$\;>$\kern-0.75em\raise-1.1ex\hbox{$\sim\;$}}}
\def\lsim{\raise0.3ex\hbox{$\;<$\kern-0.75em\raise-1.1ex\hbox{$\sim\;$}}}

\def\beqn#1{\begin{equation}\label{#1}}
\def\eeqn{\end{equation}}

\def\beqa#1{\begin{eqnarray}\label{#1}}
\def\eeqa{\end{eqnarray}}

\def\Z2{$\mathcal{Z_2}$}

\newcommand {\ignore}[1]{}

\begin{document}

\title{Stochastic inflation as an open quantum system
}

\author{Yue-Zhou Li}
\email{liyuezhou@princeton.edu}
\affiliation{Department of Physics, Princeton University, Princeton, NJ 08540, USA}

\begin{abstract} 
We reinterpret Starobinsky’s stochastic inflation as an open quantum system, where short-wavelength modes act as the environment for long-wavelength modes. Using the Schwinger–Keldysh formalism, we systematically trace out the environment and derive an effective theory for the reduced density matrix, including deviations from exact de Sitter.
The resulting master equation is a Lindblad equation, which reduces to a Fokker–Planck equation for the diagonal elements up to higher orders in the slow-roll expansion, while also yielding a more complete equation in phase space. Finally, we extend the formalism to global de Sitter, for which the associated Fokker–Planck equation lacks equilibrium solutions until the late-time regime $aH \gg 1$.

\end{abstract}

\maketitle

\textit{Introduction--} Open quantum systems are quantum systems that interact with an environment. The relevant phenomena are ubiquitous, even in daily life, as they often encode the emergence of classical stochastic behavior from quantum theory \cite{kamenev2023field,Sieberer:2015svu,Sieberer:2023qyv}. %This ubiquity arises because perfectly isolated systems are too idealized to exist in practice. 
In this philosophy, the “environment” refers to any degrees of freedom not explicitly tracked in our physical description (see, e.g., \cite{breuer2002theory,rotter2015review,de2017dynamics} for reviews of open quantum systems). %For example, focusing on the motion of a few particles in a gas while coarse-graining over the rest leads to an effective open system, which exhibits stochastic behavior in the classical limit \cite{}.

This raises a natural question: should we treat effective descriptions of gravitating systems, especially our universe, as closed or open, given that we cannot explicitly track all quantum degrees of freedom of gravity?

A first attempt to address this was made by Starobinsky in the context of inflationary cosmology \cite{Starobinsky:1986fx}. Inflation is an epoch of exponential expansion before the hot big bang in our universe, during which the inflaton slowly rolls down a quite flat potential, and its quantum fluctuations are stretched to macroscopic scales, seeding the cosmic microwave background. However, when these fluctuations become comparable to the classical slow-roll motion, standard cosmological perturbation theory  (see, e.g., \cite{Mukhanov:1990me,Maldacena:2002vr,Acquaviva:2002ud,Durrer:2004fx,Baumann:2009ds,wang2014inflation}) become unreliable, and quantum effects can push the inflaton upward, leading to eternal inflation \cite{Linde:1986fd,Goncharov:1987ir,Guth:2007ng,Creminelli:2008es}.

Starobinsky formulated this picture as a stochastic process, modeling the inflaton as undergoing a random walk governed by a Langevin equation and an associated Fokker–Planck (FP) equation \cite{Starobinsky:1986fx,Starobinsky:1994bd}. This is known as stochastic inflation, and it has been widely used to understand eternal inflation \cite{Goncharov:1987ir,mijic1990random,Linde:1993nz,Linde:1993xx,Tolley:2008na} and the resummation of secularly growing cosmological observables \cite{Starobinsky:1994bd,tsamis2005stochastic,gorbenko2019lambda,Woodard:2025cez,rey1987dynamics,nambu1988stochastic,nambu1989stochastic,kandrup1989stochastic,habib1992stochastic,Enqvist:2008kt,Baumgart:2019clc}. However, most formulations and generalizations rely on the Langevin equations and a postulated FP equation \cite{Calzetta:1993qe,PerreaultLevasseur:2013kfq,Fujita:2013cna,Fujita:2014tja,PerreaultLevasseur:2014ziv,Vennin:2015hra,Firouzjahi:2018vet,Pinol:2018euk,Firouzjahi:2020jrj,Pattison:2021oen,Cruces:2021iwq,cohen2021stochastic,Aldabergenov:2025ulq}, with only limited understandings from underlying quantum field theory (QFT) in curved space \cite{morikawa1990dissipation,hosoya1989stochastic,Collins:2017haz,Andersen:2021lii,Tokuda:2017fdh,Tokuda:2018eqs,Miyachi:2023fss,Pinol:2020cdp} (see \cite{Cruces:2022imf} for a review and references therein).

The key point is that the stochastic picture is essentially the semiclassical limit of an open quantum system \cite{kamenev2023field,Sieberer:2015svu}. A well-known example is the Brownian particle \cite{schwinger1961brownian,massignan2015quantum,maniscalco2004lindblad,lampo2016lindblad,kamenev2023field}, which is precisely the phenomenon to which stochastic inflation draws an analogy. Therefore, a natural way to understand stochastic inflation is to treat it as an open quantum system, where short-wavelength modes act as the environment for the long-wavelength modes (see \cite{Burgess:2006jn,Tolley:2008qv,Burgess:2014eoa,Salcedo:2024smn,Colas:2024lse,Shandera:2017qkg,DaddiHammou:2022itk,deKruijf:2024ufs,Lopez:2025arw} for relevant discussions). Starobinsky’s stochastic formulation then emerges as the semiclassical limit of the open quantum dynamics of the universe.

In this Letter, we develop this philosophy into a concrete and controlled framework. We treat the entire universe undergoing eternal slow-roll inflation as a quantum system, and explicitly trace out short-wavelength modes to obtain a mixed cosmological density matrix. Using the Schwinger–Keldysh (SK) path integral formalism \cite{schwinger1961brownian,keldysh2024diagram}, we derive the master equation of this density matrix. Restricting to its diagonal elements, we obtain the FP equation to higher orders in the slow-roll expansion (which recovers Starobinsky's result at leading order \cite{Starobinsky:1986fx}). Projecting to phase space (known as the Wigner function), we derive the corresponding master equation and find that, up to the order we compute, it takes the form of a Lindblad equation \cite{breuer2002theory,manzano2020short}. %This provides, to our knowledge, the first derivation of a Lindblad equation from first principles in a gravitational QFT by explicitly integrating out ultraviolet modes.

This framework also generalizes to global de Sitter (dS) space, relevant to quantum cosmology \cite{hartle1983wave,Lehners:2023yrj,Maldacena:2024uhs}, where we derive the corresponding FP equation and show that the early-time universe does not approach equilibrium states until its physical size far exceeds the Hubble radius.

In this philosophy, the FP equation and its phase space generalization should be interpreted as Lindbladian extensions of the Wheeler–DeWitt (WdW) equation \cite{dewitt1967quantum} in the minisuperspace of quantum gravity in the semi-classical limit.

\textit{Density matrix evolution from Schwinger–Keldysh formalism--} Let us briefly establish the key framework for exploring stochastic inflation as an open quantum system.

\begin{figure}[h]
    \begin{center}
    \begin{tikzpicture}[decoration={markings, mark= at position 0.52 with {\arrow{stealth}}}]
    % Draw the axes
    \node (a) at (-2,2) {$t$};
	\draw (a.north west) -- (a.south west) -- (a.south east);
    \draw[->] (-2.8, 0) -- (2.8, 0) node[right] {${\rm Re}\, t$};
   \draw[thick,postaction={decorate}] (0,2) -- (0,0.1)  node at (0.7,1) {};
   \draw[] node at (0.4,2) {$i\infty$};
   \draw[] node at (0.5,-2) {$-i\infty$};
\draw[thick,postaction={decorate}] (0,0.1) -- (2,0.1)
node at (1.2,0.35) {$x^+,{\rm ket}$}; \draw[thick,postaction={decorate}] (2,-0.1) -- (0,-0.1) ; \draw[thick,postaction={decorate}] (0,-0.1) -- (0,-2)
     node at (1.2,-0.35) {$x^-,{\rm bra}$};
 \node at (2.4,0.3) {$\rho(t)$};
\end{tikzpicture}
    \end{center}
    \caption{The SK contour prepares and evolves the density matrix from Euclidean infinity to the real time at which it is measured. In the context of gravity, the doubled variables $x^\pm$ extend to include both the quantum fields and the geometry \cite{hartle1983wave,Ivo:2024ill}.}
    \label{TimeContour}
\end{figure}
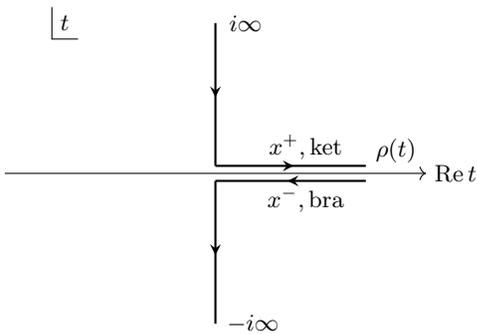

For any given theory with action $S_0[x, p]$, where $x$ is the field variable and $p$ its conjugate momentum, the density matrix of the “minimal energy” state can be prepared and evolved forward in time using the SK path integral, which connects Euclidean time to real time, as illustrated in Fig.~\ref{TimeContour}. 

The SK formalism doubles the degrees of freedom, introducing $(x^{\pm}, p^{\pm})$, and treats $\pm$ distinctly as the ket and bra components. %of the density matrix. 
The Euclidean segment selects the vacuum initial state, which in our case is the Bunch–Davies (BD) state \cite{bunch1978quantum}. The density matrix can then be projected into any representation by specifying the boundary condition at the final time $t$, at which the density matrix is evaluated.
\be
\langle\langle r[x^\pm,p^\pm]|\rho(t)\rangle\rangle=N^2 \int_{x^\pm(\pm i \infty=0)}^{r[x^\pm,p^\pm]} Dx^{\pm}Dp^{\pm}e^{iS_{L0}(x^\pm,p^\pm)}\,,
\ee
where $S_{L0}=S^+_0-S^-_0$. In this formula, we adopt the Liouville space \cite{manzano2020short}, in which the density matrix is promoted to a vector, leading to
\be
S_{L0}=\int dt \left(p^c \dot{x}^q+p^q \dot{x}^c-H_L\left(x^c,x^q,p^c,p^q\right)\right)\,,
\ee
where we introduce the Keldysh basis $x^c=\frac{x^++x^-}{2}$ the average field and $x^q=x^+-x^-$ the response field.
$H_L$ is the Liouville superoperator, which plays the role of the Hamiltonian in Liouville space. In this Letter, we consider the representation $r[x^\pm, p^\pm]$ in two scenarios: 
%(I) the diagonal elements in the position representation, and (II) the Wigner function in phase space.
\begin{itemize}
\item[I.] Diagonal density matrix $x^c(t)=x\,, x^q(t)=0$
\be
\langle \langle r|\rho(t)\rangle\rangle:=P(x,t)\,.
 \ee
\item[II.] The Wigner function $x^c(t)=x\,,p^c(t)=p$
\be
 \langle \langle r||\rho(t)\rangle\rangle=\int \frac{dx'}{\pi}\langle x+x'|\rho|x-x'\rangle e^{2ip x'}:=W(x,p,t)\,.
\ee
\end{itemize}
We now consider the case where $x$ is coupled to an environment $y$. The variable $y$ may represent an arbitrary auxiliary system; in particular, it can represent the short-wavelength modes of the same underlying system. In this context, the system–environment separation resembles the scale separation in Wilsonian effective field theories (EFTs). Tracing out the environment then corresponds to constructing an effective theory by integrating out $y$. This procedure introduces the Feynman-Vernon influence functional $e^{i S_{\rm IF}}$ \cite{feynman1963theory}, encoding imprints of system–environment interactions
\be
\langle\langle r[x^\pm,p^\pm]|{\rm Tr}_y\rho(t)\rangle\rangle=N^2 \int_{x^\pm(\pm i \infty=0)}^{r[x^\pm,p^\pm]}Dx^\pm Dp^\pm e^{i\left(S_{L0}+S_{\rm IF}\right)}\,.\label{eq: reduced density matrix}
\ee
The effective theory is $S_L^{\rm eff} = S_{L0} + S_{\rm IF}$, with its Liouville Hamiltonian $H_L^{\rm eff}$, subject to the appropriate boundary conditions. This gives the master equation for the reduced density matrix
\be
\partial_t\rho_x(r[x^\pm,p^\pm],t)=N^2 \int_{x^\pm(\pm i \infty=0)}^{r[x^\pm,p^\pm]}Dx^\pm Dp^\pm \left(-i H_{L}^{\rm eff}\right)e^{i S_{L}^{\rm eff}}\,.
\ee
This equation is not closed in general. Nevertheless, for a diagonal density matrix, it is clear that we should simply replace $p^q = -i \partial_x$, we simultaneously evaluate other variables $p^c$ using their classical solutions subject to the boundary conditions. For the Wigner function, the master equation becomes manifestly closed upon replacing $p^q = -i \partial_x$ and $x^q = i \partial_p$.
\begin{align}
& \partial_t P(x,t)=-i \left[H_L^{\rm eff}\left(x,x^q, p^c,-i\partial_x \right)\right]_{\rm saddle}P(x,t)\,,\nonumber\\
& \partial_t W(x,p,t)=-i H_L^{\rm eff}\left(x,i\partial_p,p,-i\partial_x\right) W(x,p,t)\,.\label{eq: master equations}
\end{align}

\textit{Effective theory from tracing out short wavelength environment--} We now apply this formalism to single-field inflation $S=S_{\rm grav}+S_{\phi}$ where
\begin{align}
& S_{\rm grav}=\frac{1}{16\pi G}\left(\int d^4x\sqrt{-g}R-2\int d^3x\sqrt{h}K\right)\,,\nonumber\\
& S_{\phi}=-\int d^4x\sqrt{-g}\left(\frac{1}{2}(\nabla\phi)^2+V(\phi)\right)\,.
\end{align}
%We will derive the framework of stochastic inflation by interpreting it as an open quantum system.

We consider the slow-roll regime, defined by $16\pi G\, \epsilon = (V'/V)^2 \ll 1$. Gravity is treated semi-classically after inflation starts at $\phi = \phi_0$ and $V(\phi_0) = 8\pi G H_0^2/3$, where $G H_0^2 \ll 1$. As a consequence, we work with a classical background metric $ds^2 = -dt^2 + a^2 dx^2$, extending it to SK geometry $(ds^\pm)^2$ and retaining only terms linear in the quantum component $a^q$ in the SK action $S^+ - S^-$. These terms generate classical time evolution for the scalar field via the Liouville Hamiltonian constraint
\be
\left(H_{L, \text{grav}} + H_{L, \phi}\right)\rho = 0\,, \quad H_{L, \text{grav}} \simeq -i \dot{a}^c \partial_{a^c} = -i\partial_t\,.
\ee
We focus on the SK action of the scalar field in a fixed classical background $a^c=a_0^c\left(1+\cdots\right)\,, a_0^c=e^{H_0 t}$, with possible corrections to the exact dS (we slip off the superscript from now on). A detailed treatment is provided in the Supplemental Material.

We nevertheless allow for eternal inflation with $H_0^2/\epsilon \sim M_{\rm pl}^2$. In this regime, quantum fluctuations dominate and standard perturbation theory fails. However, since the background remains classical and adiabatically evolving, we can perform a controlled expansion in $|V - V_0| / V_0 \sim \sqrt{\epsilon}\, \Delta\phi\ll 1$. The original picture is to think of quantum effects as noise inducing a random walk of the inflationary trajectory \cite{Starobinsky:1986fx}. In our paradigm, this picture is made precise by treating the long modes as an open system that defines the trajectory, with short modes as its environment: $\phi = \phi_s + \phi_e$, where $\phi_e(k,t) = \theta(k - \varepsilon a_0 H_0)\, \phi(k,t)$. The step function separation enforces the Markovian approximation.\footnote{Other coarse-graining functions also exist, introducing colored noise \cite{Winitzki:1999ve}, which gives rise to non-Markovian memory effects. Nevertheless, as long as the separation function approaches a step function controlled by some parameter $b$, the Markovian approximation can be recovered by taking its series expansion in $b$.} We take the infrared parameter $\varepsilon \to 0$ \cite{Starobinsky:1986fx}, so that $\phi_s$ captures zero modes. This realizes a scale separation analogous to Wilsonian EFT, though it goes beyond the standard Wilsonian treatment due to the time dependence of the cutoff $\varepsilon a_0 H_0$.

The reduced density matrix of $\phi_s$ is then computed by tracing out the short modes as the environment, replacing $(x,p)\rightarrow (\phi_s,\Pi_s)$ and $y\rightarrow \phi_e$ in \eqref{eq: reduced density matrix}.
We have
\begin{widetext}
\begin{align}
& S_{L0}=\int dt d^3x\left(a^3 \dot{\phi}_s^q\dot{\phi}_s^c-a^3 v'(\phi_s^c)\phi_s^q-a \partial_i\phi_s^q \partial^i\phi_s^c+\mathcal{O}\left((\phi_s^q)^3\right)\right)\,,\quad e^{i S_{\rm IF}}=\int_{\rm BD}D\phi_e^\pm e^{i\left(S_{Le0}+S_{\rm int}\right)} \nonumber\\
& S_{Le0}=\int dtd^3x\left(a^3\dot{\phi}_e^q \dot{\phi}_e^c-a\partial_i\phi_e^q \partial^i\phi_e^c\right)\,,\quad S_{\rm int}=\int dtd^3x\left(a^3\dot{\phi}^q_{(s}\dot{\phi}^c_{e)}-a\partial_i\phi^q_{(s}\partial^i\phi^c_{e)}+\delta v_{\rm int}\right)\,,
\end{align}
\end{widetext}
where $\delta v_{\rm int}=v(\phi_s^+)-v(\phi_s^-)-v(\phi^+)+v(\phi^-)$ with $v=V-V_0$. In this Letter, we only retain $\phi_s^q$ terms up to quadratic order, capturing the leading effects of diffusion and decoherence.

Tracing out the environment leads to an effective theory to order $\mathcal{O}(v^2,(\phi^q_s)^2,\partial_t^3,k^2)$, with $S_{L}^{\rm eff}=S_{Lr}+S_{\rm IF}$ where\footnote{We retain only the zero modes, $\phi_s \simeq \phi_{s0} + \mathcal{O}(k)$, and regulate the comoving volume $(2\pi)^3 \delta^{(3)}(0)$ as $\Omega$. For convenience, we drop the subscript $0$ and work in a unit comoving volume, $\Omega = 1$, so that the zero modes are canonically normalized. Equivalently, we can also think of this unit as rescaling $\phi^q \rightarrow \phi^q/\Omega$. Therefore, the choice $\Omega = 1$ is independent of the unit convention $8\pi G = 1$, and at most alters the normalization convention of retarded correlators of zero modes.}
\begin{widetext}
\begin{align}
& S_{Lr}=\int dt\left(a^3 \dot{\phi}_s^q \dot{\phi}^c_s- a^3 v_r'(\phi_s^c)\phi_s^q+\mathcal{O}\left((\phi_s^q)^3,k^2)\right)\right)\,,\nonumber\\
& S_{\rm IF}=i\int dt \, a^6\left(C_1(\phi^c_s)\left(\dot{\phi}^q_s\right)^2+C_2(\phi^c_s)\dot{\phi}^q_s \phi^q_s+C_3(\phi^c_s)\left(\phi^q_s\right)^2+\mathcal{O}\left(\left(\phi^q_s\right)^3,k^2,\partial_t^3,\delta C_{ir} \right)\right)\,.\label{eq: EFT}
\end{align}
\end{widetext}
The diffusion coefficients $C_i$ are determined by Feynman diagrams involving virtual short modes\footnote{This structure is known as the open EFT of cosmology \cite{LopezNacir:2011kk,Burgess:2014eoa,Salcedo:2024smn,Colas:2024lse,Salcedo:2025ezu}. It also suggests that the universe can be modeled as hydrodynamical system by following the logic of \cite{Tasinato:2025zqt,Zhou:2025mbq}.}. $v_r, \delta C_{ir}$ refer to the renormalization effects. The new operators $(\dot{\phi}^q_s)^2, \dot{\phi}^q_s \phi^q_s, (\phi^q_s)^2$ encode the diffusion in the phase space. More importantly, $(\phi_s^q)^2$ term is responsible for the decoherence in the pointer basis $\phi^\pm$. The system undergoes the decoherence in $\phi^\pm$ basis as far as $C_3>0$ , allowing for possible quantum-to-classical transition \cite{morikawa1987origin,morikawa1990dissipation,Lombardo:1995fg,Burgess:2006jn,Burgess:2014eoa,Sano:2025ird,Sano:2025ird,Liu:2016aaf}. However, the actual quantum-to-classical transition should be captured by the phase space decoherence, where it is relevant to $\mathcal{C}={\rm det}\left(
\begin{array}{cc}
 2 C_1 & C_2 \\
 C_2 & 2 C_3 \\
\end{array}
\right)$.

\textit{1. Assumptions and Feynman rules--} We now detail the approximations entering the EFT \eqref{eq: EFT}. we consider two standard assumptions commonly used in open quantum systems \cite{manzano2020short}: (1) the Born approximation, and (2) the secular approximation. The Born approximation assumes weak system–environment coupling, allowing a perturbative expansion in terms of Feynman diagrams. The secular approximation suppresses fast-oscillating terms relative to the system’s time scale and corresponds to large time-scale separation in the Wilsonian EFT sense. This separation leads to a multipole expansion in time, e.g., $\phi_s(t_2) = \phi_s(t_1) + \partial_{t_1} \phi_s(t_1)(t_2 - t_1) + \cdots$, such that the system field acts as an effectively static source of environment and does not enter the $t_2$ integral. The Markovian approximation is perturbatively ensured by the step-function separation between the system and the environment, since it projects the environment’s response to equal times, which vanishes in the BD vacuum, $G_R(t,t) = 0$. A similar scale separation exists in spatial momentum $k$, enabling a multipole expansion in space as well.

We summarize here the relevant Feynman rules used to trace out $\phi_e$ (with conformal time $\eta H_0=-e^{-H_0 t}$).
We use vertices
\begin{align}
& \begin{tikzpicture}[baseline={(0,-0.5ex)}]
\draw[thick,dashed] (0,0) -- (0.7,0);
  \draw[thick] (0.7,0) -- (1.4,0);
  \node at (0.6,0.2) {$q$};
  \node at (0.8,0.2) {$c$};
  \node at (0.7,-0.3) {$t$};
\end{tikzpicture}=a^3 \dot{\phi}_s^q \partial_t-a k^2 \phi_s^q\,,\quad
\begin{tikzpicture}[baseline={(0,-0.5ex)}]
  \draw[thick] (0,0) -- (0.6,0);
   \draw[thick] (0.8,0) -- (1.4,0);
  \draw[thick] (0.7,0) circle (0.1);
   \draw[thick] (0.7-0.07, 0-0.07) -- (0.7+0.07, 0+0.07);
  \draw[thick] (0.7-0.07, 0+0.07) -- (0.7+0.07, 0-0.07);
  \node at (0.5,0.2) {$c$};
  \node at (0.9,0.2) {$q$};
  \draw[thick,dashed] (0.7,0.1) -- (0.7,0.6);
  \node at (0.7,0.7) {$c$};
  \node at (0.7,-0.3) {$t$};
\end{tikzpicture}= v''(\phi_s^c(t))\,.
\end{align}
The propagator is
\begin{widetext}
\begin{align}
&
\begin{tikzpicture}[baseline={(0,-0.5ex)}]
  \draw[thick] (0,0) -- (1.4,0);
  %\draw[thick,->] (0,0) -- (0.75,0);
  \node at (0.1,0.2) {$c$};
  \node at (1.3,0.2) {$c$};
  \node at (0.1,-0.2) {$t_1$};
  \node at (1.3,-0.2) {$t_2$};
  \node at (0.75,0.25) {$k$};
\end{tikzpicture}=\frac{H_0^2 \left(\left(\eta _1 \eta _2 k^2+1\right) \cos \left(\left(\eta _1-\eta _2\right) k\right)+\left(\eta _1-\eta _2\right) k \sin \left(\left(\eta _1-\eta _2\right) k\right)\right)}{2k^3}\theta\left(k+\frac{\varepsilon}{\eta_1}\right)\theta\left(k+\frac{\varepsilon}{\eta_2}\right)\,,\nonumber\\
& \begin{tikzpicture}[baseline={(0,-0.5ex)}]
  \draw[thick] (0,0) -- (1.4,0);
  %\draw[thick,->] (0,0) -- (0.75,0);
  \node at (0.1,0.2) {$c$};
  \node at (1.3,0.2) {$q$};
  \node at (0.1,-0.2) {$t_1$};
  \node at (1.3,-0.2) {$t_2$};
   \node at (0.75,0.25) {$k$};
\end{tikzpicture}=-i\frac{H_0^2\left(\left(\eta _1 \eta _2 k^2+1\right) \sin \left(\left(\eta _1-\eta _2\right) k\right)+\left(\eta _2-\eta _1\right) k \cos \left(\left(\eta _1-\eta _2\right) k\right)\right)}{k^3}\theta\left(t_1-t_2\right)\theta\left(k+\frac{\varepsilon}{\eta_1}\right)\theta\left(k+\frac{\varepsilon}{\eta_2}\right)\,.\label{eq: propagators}
\end{align}
\end{widetext}
We use solid line for $\phi_e$ and dashed line for $\phi_s$. Similar Feynman rules are also used in \cite{Tokuda:2017fdh,Tokuda:2018eqs}.

\textit{2. Feynman diagrams and diffusion coefficients--} For example, if we consider an exact dS $a_0=e^{H_0 t}$, the low-lying relevant diagrams that generate the diffusion up to $v^2$ are:
\begin{widetext}
\be
\begin{tikzpicture}[baseline={(0,-0.5ex)}]
\draw[thick,dashed] (-0.5,0.5)--(0,0);
\draw[thick] (0,0)--(1,0);
\draw[thick,->] (0,0)--(0.6,0);
\draw[thick,dashed] (1,0)--(1.5,0.5);
\node at (-0.05,0.25) {$q$};
\node at (1+0.05,0.25) {$q$};
\node at (0.15,0.15) {$c$};
\node at (0.85,0.15) {$c$};
\node at (0,-0.3) {$t_1$};
\node at (1,-0.3) {$t_2$};
\end{tikzpicture}+2\times
\begin{tikzpicture}[baseline={(0,-0.5ex)}]
\draw[thick,dashed] (-0.5,0.5)--(0,0);
\draw[thick] (0,0)--(0.65,0);
\draw[thick,->] (0,0)--(0.45,0);
\draw[thick,->] (0.85,0)--(1.25,0);
\draw[thick] (0.85,0)--(1.5,0);
\draw[thick] (0.75,0) circle (0.1);
   \draw[thick] (0.75-0.07, 0-0.07) -- (0.75+0.07, 0+0.07);
  \draw[thick] (0.75-0.07, 0+0.07) -- (0.75+0.07, 0-0.07);
\draw[thick,dashed] (1.5,0)--(2,0.5);
\draw[thick,dashed] (0.75,0.1) -- (0.75,0.5);
\node at (-0.05,0.25) {$q$};
\node at (1+0.05+0.5,0.25) {$q$};
\node at (0.15,0.15) {$c$};
\node at (0.85+0.5,0.15) {$c$};
\node at (0,-0.3) {$t_1$};
\node at (1+0.5,-0.3) {$t_2$};
\node at (0.75,0.6) {$c$};
\node at (0.75,-0.3) {$t_3$};
\node at (0.55,0.15) {$c$};
\node at (0.95,0.15) {$q$};
\end{tikzpicture}+2\times
\begin{tikzpicture}[baseline={(0,-0.5ex)}]
%\draw[thick,->] (0,0)--(0.45,0);
%\draw[thick,->] (0.85,0)--(1.25,0);
%\draw[thick,->] (1.6,0) --(2,0);
\draw[thick,dashed] (-0.5,0.5)--(0,0);
\draw[thick] (0,0)--(0.65,0);
\draw[thick] (0.85,0)--(1.4,0);
\draw[thick] (1.6,0)--(2.25,0);
\draw[thick] (0.75,0) circle (0.1);
   \draw[thick] (0.75-0.07, 0-0.07) -- (0.75+0.07, 0+0.07);
  \draw[thick] (0.75-0.07, 0+0.07) -- (0.75+0.07, 0-0.07);
\draw[thick,dashed] (2.25,0)--(2.75,0.5);
\draw[thick,dashed] (0.75,0.1) -- (0.75,0.5);
\draw[thick] (1.5,0) circle (0.1);
   \draw[thick] (1.5-0.07, 0-0.07) -- (1.5+0.07, 0+0.07);
  \draw[thick] (1.5-0.07, 0+0.07) -- (1.5+0.07, 0-0.07);
\draw[thick,dashed] (2.25,0)--(2.75,0.5);
\draw[thick,dashed] (0.75,0.1) -- (0.75,0.5);
\draw[thick,dashed] (1.5,0.1) -- (1.5,0.5);
\node at (-0.05,0.25) {$q$};
\node at (1+0.05+0.5+0.75,0.25) {$q$};
\node at (0.15,0.15) {$c$};
\node at (0.85+0.5+0.75,0.15) {$c$};
\node at (0,-0.3) {$t_1$};
\node at (1+0.5+0.75,-0.3) {$t_2$};
\node at (0.75,0.6) {$c$};
\node at (1.5,0.6) {$c$};
\node at (0.75,-0.3) {$t_3$};
\node at (1.5,-0.3) {$t_4$};
\node at (0.55,0.15) {$c$};
\node at (0.95,0.15) {$q$};
\node at (1.3,0.15) {$c$};
\node at (1.7,0.15) {$q$};
\end{tikzpicture}+
\begin{tikzpicture}[baseline={(0,-0.5ex)}]
%\draw[thick,->] (0,0)--(0.45,0);
%\draw[thick,->] (0.85,0)--(1.25,0);
%\draw[thick,->] (1.6,0) --(2,0);
\draw[thick,dashed] (-0.5,0.5)--(0,0);
\draw[thick] (0,0)--(0.65,0);
\draw[thick] (0.85,0)--(1.4,0);
\draw[thick] (1.6,0)--(2.25,0);
\draw[thick] (0.75,0) circle (0.1);
   \draw[thick] (0.75-0.07, 0-0.07) -- (0.75+0.07, 0+0.07);
  \draw[thick] (0.75-0.07, 0+0.07) -- (0.75+0.07, 0-0.07);
\draw[thick,dashed] (2.25,0)--(2.75,0.5);
\draw[thick,dashed] (0.75,0.1) -- (0.75,0.5);
\draw[thick] (1.5,0) circle (0.1);
   \draw[thick] (1.5-0.07, 0-0.07) -- (1.5+0.07, 0+0.07);
  \draw[thick] (1.5-0.07, 0+0.07) -- (1.5+0.07, 0-0.07);
\draw[thick,dashed] (2.25,0)--(2.75,0.5);
\draw[thick,dashed] (0.75,0.1) -- (0.75,0.5);
\draw[thick,dashed] (1.5,0.1) -- (1.5,0.5);
\node at (-0.05,0.25) {$q$};
\node at (1+0.05+0.5+0.75,0.25) {$q$};
\node at (0.15,0.15) {$c$};
\node at (0.85+0.5+0.75,0.15) {$c$};
\node at (0,-0.3) {$t_1$};
\node at (1+0.5+0.75,-0.3) {$t_2$};
\node at (0.75,0.6) {$c$};
\node at (1.5,0.6) {$c$};
\node at (0.75,-0.3) {$t_3$};
\node at (1.5,-0.3) {$t_4$};
\node at (0.55,0.15) {$q$};
\node at (0.95,0.15) {$c$};
\node at (1.3,0.15) {$c$};
\node at (1.7,0.15) {$q$};
\end{tikzpicture}\label{eq: Feynman diagrams}
\ee
\end{widetext}
The intermediate times $(t_3, t_4)$ are integrated using a multipole expansion, e.g., $v(\phi^c(t_3)) \sim v(\phi^c(t_1)) + \cdots$. At leading order, the scale separation in wavelength sets all momenta equal, $k_1 = k_2 = \cdots$, so the “potential” insertions act as static sources. %generating loop diagrams in time. 
We retain only the leading term in both temporal and spatial multipole expansions, as higher-order terms correspond to higher-dimensional operators in EFT that are not considered in this Letter. For example, the first diagram gives the standard noise contribution $H_0^3/(4\pi^2)$ \cite{Starobinsky:1986fx}. We leave the details of this computation in the Supplemental Material.

In addition, many other diagrams exist, which are systematically captured by the standard Wilsonian coarse-graining. These renormalizing the potential $v(\phi^c)\rightarrow v_r(\phi^c)$ and the diffusion terms, and generate a tower of higher-dimensional operators in EFT. We neglect these terms (but keep the superscript $v_r(\phi_s^c)$), which corresponds to determining the relevant ``tree-level'' Wilson coefficients, including those for diffusion, by matching relevant diagrams in the full theory \cite{Salcedo:2024smn,Colas:2024lse}.

The diffusion coefficients are
\begin{widetext}
\begin{align}
& C_1(\phi^c_s)=\frac{H_0^3}{8\pi^2}\left(1+\frac{2}{3H_0^2}v''(\phi^c_s) \log+ \frac{2}{9H_0^4}v''(\phi^c_s)^2\left(\log^2+\frac{\log}{3}+\frac{\pi^2-8}{4}\right) \right)\,,\nonumber\\
& C_2(\phi^c_s)=\frac{H_0^2}{12\pi^2}v''(\phi^c_s)+\frac{1}{18\pi^2}v''(\phi^c_s)^2\left(\log+\frac{1}{6}\right)\,,\quad C_3(\phi^c_s)=\frac{H_0}{72\pi^2}v''(\phi^s_c)^2\,,\label{eq: Wilson coefficients}
\end{align}
\end{widetext}
where 
$\log=\log\frac{\varepsilon}{2}-\psi\left(\frac{3}{2}\right)$. $C_i$s can also be obtained by treating the short modes as effectively massive, with mass $\sim v''(\phi_s^c)$, as shown in the Supplemental Material. Nevertheless, we see $\mathcal{C}=0$, indicating there is no decoherence in phase space at this order\footnote{Nevertheless, we emphasize that the paradigm of open quantum systems is insightful for understanding the decoherence of the early universe, either within or beyond the Markovian limit, see, e.g., \cite{Lombardo:2004fr,Lombardo:2005iz,Burgess:2014eoa,DaddiHammou:2022itk,deKruijf:2024ufs,Lopez:2025arw} for explorations. Our formalism is thus expected to provide a systematic way of exploring and clarifying the decoherence structure, which we leave for future studies.}.

We can also include the classical effect of deviation from exact dS, with the scale factor $a = e^{H_0 t} \left(1 + \frac{8\pi G}{6H_0^2}\int^{t} dt'  \delta(\phi_s^c(t'))\right)$ such that $H=H_0+\fft{8\pi G}{6H_0}\delta(\phi^c_s)$. This deviation is suppressed in $G \to 0$ and the function $\delta(\phi_s^c)$ depends on the inflationary trajectory. Under our approximations, the deviation generates new Feynman diagrams at leading order in $\delta$ and leads to a shift in the diffusion coefficient (setting $8\pi G = 1$) $v''(\phi_s^c) \to v''(\phi_s^c) - \frac{3}{4} \delta(\phi_s^c)$ up to $\mathcal{O}(v \delta)$. See the Supplemental Material for details.

%Finally, we emphasize that our formalism can be systematically extended to construct the long-wavelength EFT, including higher-dimensional operators, effects from small but finite $(k, \varepsilon)$, and the $\beta$-functions of Wilson coefficients.

\textit{Master equations and the Lindbladian--} We can now apply \eqref{eq: master equations} to derive the master equations for stochastic inflation \eqref{eq: EFT} subject to its environment. We find 
\begin{align}
&H_L^{\rm eff}=\frac{\Pi^c \Pi^q}{a^3}+a^3 \phi^q v_r'(\phi^c)\nonumber\\
&-i\left(C_1(\phi^c)(\Pi^q)^2+a^3 C_2(\phi^c)\Pi^q\phi^q+a^6 C_3(\phi^c)(\phi^q)^2\right)\,.\label{eq: Hamiltonian L}
\end{align}

\textit{1. Fokker-Planck equation--} For diagonal density matrix, we evaluate the Hamiltonian at the saddle point on the time surface where we evaluate the density matrix, including the deviation from exact dS. In the slow-roll limit %and $v\rightarrow 0$ limit, 
we find $\phi^q_{\rm sad}=\Pi^q_{\rm sad}=0, \delta=v_r+\mathcal{O}(v_r^2)$ and
\be
 \Pi^c_{\rm sad}=-\frac{a^3}{3H(\phi^c)}\left(v_r'(\phi^c)+\frac{v_r'(\phi^c)v_r''(\phi^c)}{\left(3H(\phi^c)\right)^2}+\mathcal{O}(v^3)\right)\,,\label{eq: saddle momentum}
\ee
where $ H(\phi^c)=H_0+\frac{\delta(\phi^c)}{6H_0}+\mathcal{O}(v^2)$.
Applying \eqref{eq: master equations} leads to a Fokker-Planck equation\footnote{See \cite{Cespedes:2023aal} for a derivation of the leading FP equation from a holographic renormalization perspective, where late-time evolution is interpreted as boundary renormalization \cite{Heemskerk:2010hk}. As in effective field theories more generally, this renormalization structure implements the resummation of IR logarithms \cite{Baumgart:2019clc}.}\footnote{See \cite{Pinol:2018euk,Pinol:2020cdp} for the generalization of FP equation to multifield inflation.}
\be
\partial_t P(\phi,t)=-\partial_\phi\left(a^{-3}\Pi^c_{\rm sad}P(\phi,t)\right)+\partial_\phi^2\left(\tilde{C}_1(\phi)P(\phi,t)\right)\,,\label{eq: FP}
\ee
where $\tilde{C}_1$ is the modified diffusion coefficient by $\delta(\phi)$.
The linear $v$ in \eqref{eq: FP} reproduces the original Starobinsky FP equation \cite{Starobinsky:1986fx}. Setting $v=\lambda/4\,  \phi^4$ in exact dS $\delta=0$ in \eqref{eq: FP} reproduces the FP equation derived in \cite{gorbenko2019lambda}\footnote{To match with \cite{gorbenko2019lambda}, we need to include the renormalization that generates an effective mass $v_r(\phi)=v(\phi)+m_{\rm eff}^2 \phi^2$ and then drop higher order terms in $\lambda, \lambda m_{\rm eff}^2,\cdots$.}. We present the equilibrium states and the one-point function $\langle\phi^2\rangle$ for different potentials in Supplemental Material.

%{\color{red}(comment on the eternal inflation bound?)}

\textit{2. Wigner function and the Lindblad equation--} We now consider the Wigner function, which encodes the decoherence\footnote{See \cite{Fumagalli:2024msi} for application of the Wigner function in quantum cosmology.}.

 We first consider exact dS. Using \eqref{eq: master equations} and \eqref{eq: Hamiltonian L} yields
\begin{widetext}
\begin{align}
\partial_t W(\phi,\Pi,t)&=-a^{-3}\Pi\partial_\phi W(\phi,\Pi,t)+a^3 v_r'(\phi)\partial_\Pi W(\phi,\Pi,t)\nonumber\\
&+\partial_\phi^2\left(C_1(\phi)W(\phi,\Pi,t)\right)-a^3 \partial_\phi\partial_\Pi\left(C_2(\phi)W(\phi,\Pi,t)\right)+a^6 \partial_\Pi^2\left(C_3(\phi)W(\phi,\Pi,t)\right)\,.\label{eq: Klein-Kramers}
\end{align}
\end{widetext}
We show in Supplemental Material that \eqref{eq: Klein-Kramers} implies FP equation \eqref{eq: FP}.

The equation \eqref{eq: Klein-Kramers} is essentially a Lindblad equation for the long-wavelength inflaton. %To make this manifest, we adopt the power-counting scheme $v/H_0^2 \ll 1$ so that we can retain only the leading term in each $C_i$. 
We can show \eqref{eq: Klein-Kramers} is equivalent to projecting the Lindblad equation
$\partial_t\rho=-i[H_{r},\rho]+\gamma\left(L\rho L-\frac{1}{2}\{L^2,\rho\}\right)=-i H_{Lr}\rho-\fft{\gamma}{2} (L^q)^2 \rho$ into Wigner representation with the jump operator
\begin{align}
& L=(1+\alpha)\Pi+\frac{a^3 v''(\phi^c)(1+\beta)}{3H_0}\phi\,,\quad \gamma=\frac{H_0^3}{4\pi^2}\,,\nonumber\\
& \alpha=\frac{v''(\phi^c)}{3H_0^2}\log+\fft{v''(\phi^c)^2}{108H_0^4}\left(6\log^2+4\log+3\pi^2-24\right)\,,\nonumber\\
& \beta=\frac{v''(\phi^c)}{3H_0^2}\left(1+3\log\right)\,.
\end{align} 
Our method can nevertheless systematically go beyond the Lindbladian structure. Again, including the deviation from dS shifts $v''\rightarrow v''-3/4 \delta$ up to $\mathcal{O}(v \sigma)$ with $\delta=v_r$, as shown in the Supplemental Material.

%Therefore, we show that the Lindblad structure arises naturally from environment loops under a time-domain multipole expansion — and that our framework allows corrections beyond Lindbladian dynamics

\textit{Generalization to global slicing--} In the previous sections, our discussion has been restricted to the flat slicing, as is standard in stochastic inflation \cite{Starobinsky:1986fx}. However, the open-system paradigm and our framework can be systematically generalized to other coordinates, such as the global slicing relevant in quantum cosmology\footnote{See also \cite{Mirbabayi:2019qtx} for a discussion for the static patch.}. This allows curvature fluctuations large enough to create a sphere $\Omega_3$, as in the metric $ds^2 = -d\tau^2 + \frac{a^2}{H_0^2} d\Omega_3^2$, with $a = \cos(H_0 \tau)$ for exact dS. This setup captures the physics of earlier times $\phi_0$, when the number of e-folds is not yet large enough to flatten the spatial geometry.

In the global slicing, the field $\phi$ is decomposed into spherical harmonics as $\phi = \sum_{J m_1 m_2} \phi_\ell Y_{J m_1 m_2}(\Omega)$. The canonical modes and their SK correlators are presented in the Supplemental Material. In this setting, short modes are defined by $J > \varepsilon a_0 H_0 \gg 1$, and we trace them out in the limit $\varepsilon \ll 1$. Our analysis shows that the resulting diffusion coefficients are consistent with the expectation that short modes remain locally correlated by flat-slice approximation, although the spatial geometry remains as a sphere (see Supplemental Material). For example, at leading order we have $C_1(\phi^c_s,\tau)=\frac{H_0^3}{8\pi^2} {\rm tanh}\left(H_0\tau\right)$.

We now present the FP equation in global slicing of exact dS at leading order, while leaving a more complete analysis to the future work. This is an equation for global diagonal Hartle-Hawking density matrix \cite{hartle1983wave,Ivo:2024ill} in minisuperspace $P(\phi,a)$ 
\begin{align}
& \sqrt{H_0^2 a^2-1}\, \partial_a P(\phi,a)\nonumber\\
&=\partial_\phi\left(\frac{a \,v_r'(\phi)P(\phi,a)}{3\sqrt{H_0^2 a^2-1}}\right)+\partial_\phi^2\left(\frac{H_0^2 \sqrt{H_0^2 a^2-1}}{8\pi^2 a}P(\phi,a)\right)\,.\label{eq: FP global}
\end{align}
In the late-time limit $a H_0 \gg 1$, Eq.~\eqref{eq: FP global} reduces to the flat-slice FP equation. Nevertheless, it also shows that the early universe is not in an equilibrium state $\partial_a P=0$ prior to late times, because of the nontrivial $a$-dependence in diffusion. Conceptually, \eqref{eq: FP global} should be interpreted as semi-classical limit of a Lindblad-type extension of the WdW equation \cite{dewitt1967quantum} in the minisuperspace\footnote{Clarifying this interpretation may require a deeper understanding of the Hilbert space in quantum cosmology. See, e.g., \cite{Chandrasekaran:2022cip,Witten:2023xze,Chen:2024rpx,Kudler-Flam:2024psh,Kudler-Flam:2025pol,Harlow:2025pvj} for recent progress.}.

%Conceptually, Eq.~\eqref{eq: Klein-Kramers} can be interpreted as the semi-classical limit of a Lindblad-type extension of the Wheeler–DeWitt equation.

\textit{Conclusions.--} In this Letter, we clarify that stochastic inflation is naturally understood as an open quantum system, with short-wavelength modes acting as the environment. We develop a systematic framework to study its reduced density matrix using the SK formalism. The approximations involved in tracing out the environment are made explicit and can be viewed as a generalization of Wilsonian coarse-graining with a time-dependent scale. We then present a diagrammatic rule for computing the effective theory governing the reduced density matrix, focusing on diffusion terms and allowing for deviations from exact dS. We obtain the corresponding master equation, yielding the FP equation to higher orders in the slow-roll expansion for the diagonal elements, and its phase-space generalization for the Wigner function, which we identify as a Lindblad equation. Finally, we generalize this framework to global dS and derive the corresponding FP equation, which intrinsically does not have an equilibrium state until very late times.

Our analysis confirms that gravitational effective theories can be non-unitary, describing open quantum systems, even if their short-distance completions are closed\footnote{Similar conclusions can be reached in the context of gravitational wave physics, where the EFTs describing wave emission and tidal deformations are also open systems \cite{Goldberger:2009qd,Saketh:2023bul,Ivanov:2024sds,Glazer:2024eyi,Caron-Huot:2025tlq}.}. On the other hand, our work also illustrates that EFT provides a powerful framework for open quantum systems, which could in principle include RG \cite{Baidya:2017eho,Kim:2024rcg} and systematic corrections beyond the Lindbladian.

The explicit Lindblad structure we obtain for inflationary cosmology opens the possibility of simulating early-universe using engineered open quantum systems, such as cold atom platforms \cite{bloch2008many,bloch2012quantum}\footnote{See also \cite{Liu:2020wtr} for a quantum algorithm in the inflation context.}.

The results of this Letter nevertheless are restricted to zero modes and low-lying diffusion. The analysis of including the deviation from dS is also limited to the leading order.

In the future, it is important to systematically  tracking the renormalization and  higher-dimensional operators in EFT, including the deviation from dS to high orders. It is also interesting to explore other gauges \cite{Cruces:2021iwq} and include tensor fluctuations \cite{Pattison:2017mbe}. It will also be important to solve those higher-order FP equation and its phase space generalization to understand cosmological observables.

We emphasize that our analysis, particularly the generalization to global dS, initiates a program to find Lindblad equations in quantum gravity, generalizing WdW equation. This may offer new insights into the measure problem in eternal inflation \cite{Guth:2007ng,Winitzki:2006rn}.

\textit{Acknowledgments--} We are grateful to Thomas Colas, Sebastián Céspedes, Lennard Dufner, Victor Gorbenko, Aidan Herderschee, Victor Ivo, Juan Maldacena, Jajie Mei, Sebastian Mizera, Enrico Pajer, Shinsei Ryu, David Simmons-Duffin, Piotr Tourkine, and Zihan Zhou for useful discussions. We are also grateful to Simon Caron‑Huot, Cheng Shang, and Junsei Tokuda for their valuable comments and suggestions on the draft. The work of Y.Z.L is supported in part by the US National Science Foundation under Grant No. PHY- 2209997, and in part by Simons Foundation grant No. 917464.

\bibliography{short.bib}

\newpage 

%%%%%%%%%% Merge with supplemental materials %%%%%%%%%%
\pagebreak
\widetext
\begin{center}
\textbf{\large Supplemental Material}
\end{center}

\section{Emergence of time in semi-classical gravity}

In this material, we provide a more detailed discussion of the emergence of time in semi-classical gravity using the Schwinger–Keldysh (SK) formalism.

We consider slow-roll inflationary model $S=S_{\rm grav}+S_{\phi}$
\be
S_{\rm grav}=\frac{1}{16\pi G}\left(\int d^4x\sqrt{-g}R-2\int d^3x\sqrt{h}K\right)\,,\quad
 S_{\phi}=-\int d^4x\sqrt{-g}\left(\frac{1}{2}(\nabla\phi)^2+V(\phi)\right)\,.
\ee
For simplicity, we consider a gauge where we fix the classical background to be homogeneous
\be
ds^2=-N^2 dt^2+a^2 dx^2\,.
\ee
In this background, we have the action
\be
S=-3 \int dtd^3x \frac{a \dot{a}^2}{N}-\int dtd^3x\left(\fft{1}{2N}a^3 \dot{\phi}^2-\fft{1}{2}N a (\partial\phi)^2-a^3 N V(\phi)\right)\,.
\ee
We now explicitly write down its SK action $S_L=S_{L,{\rm grav}}+S_{L,\phi}+S_{L,{\rm res}}$ by treating gravity semi-classically to keep only linear terms in $g_{\mu\nu}^q$ and take $N^c=1$:
\begin{align}
S_{L,{\rm grav}}&=-3 \int dt d^3x \dot{a^c}\left(a^q \dot{a}^c+a^c\left(2\dot{a}^q-N^q \dot{a}^c\right)\right)\,,\quad  S_{L,\phi}=\int dt d^3x\left((a^c)^3 \dot{\phi}^c \dot{\phi}^q-a^c \partial_i \phi^c \partial^i \phi^q+(a^c)^3\left(v^+-v^-\right)\right)\,,\nonumber\\
 S_{L,{\rm res}}&=-\fft{1}{4}\int dt d^3x a^q\left(\left((\partial\phi^+)^2-3(a^c)^2 (\dot{\phi}^+)^2+6 (a^c)^2 V^+\right)+(\phi^+\rightarrow \phi^-)\right)\nonumber\\
& -\fft{1}{4}\int dt d^3x a^c N^q\left(\left((\partial\phi^+)^2+(a^c)^2 (\dot{\phi}^+)^2+2(a^c)^2 V^+\right)+\left(\phi^+\rightarrow \phi^-\right)\right)\,,
\end{align}
where $v=V(\phi)-V(\phi_0)$ and $V^\pm=V(\phi^\pm)$.
The last term $S_{L,{\rm res}}$ refers to the linear response of the metric, encoding $(a^q,N^q)$. The physical picture of having $a^q$ is that the no-boundary Hartle-Hawking geometry \cite{hartle1983wave} bends to Euclidean region differently for bra and ket and the resulting manifold is complex \cite{Witten:2021nzp}, known as the SK geometry \cite{Jana:2020vyx}. See \cite{Ivo:2024ill} for an illustration. 

The Hamiltonian constraint for the Liouvelle superoperator is
\be
\left(H_{L,{\rm grav}}+H_{L,\phi}+H_{L,{\rm res}}\right)\rho(a^\pm,\phi^\pm)=0\,,
\ee
where the Liouvelle Hamiltonian for the semi-classical gravity is:
\be
H_{L,{\rm grav}}=-\frac{p^c_a p^q_a}{6a^c}\rightarrow H_{L,{\rm grav}}\rho=-i\dot{a}^c \partial_{a^c}\rho\,.
\ee
The classical time flow for the inflaton is then generated by the operator $H a\partial_a=\dot{a}\partial_a:= \partial_t$.

We further separate $\phi = \phi_s + \phi_e$ and trace out $\phi_e$. In our approximation, however, we may neglect effects from $S_{L,{\rm res}}$. Tracing out $\phi_e$ in $S_{L,{\rm res}}$ generates quadratic diffusion terms of the form $\mathcal{O}\left((a^q)^2, a^q N^q, a^q \phi^q, N^q \phi^q\right)$, which encode quantum effects of gravity and are neglected in our approximation. Therefore, we can set $\phi = \phi_s$ in $S_{L,{\rm res}}$ and focus on tracing out $\phi_e$ solely within $S_{L,\phi}$, with a fixed classical scale factor $a^c$, as discussed in the main text. Nevertheless, we emphasize that the classical scale factor $a^c$ should be determined by varying $(a^q, N^q)$ in $S_{L,{\rm grav}} + S_{L,{\rm res}}$, which yields the Friedmann equations. At leading order around $\phi=\phi_0$, these equations give $a^c=e^{H_0 t}$.

\section{Computations of diffusion coefficients}

In this material, we provide more details to deriving the diffusion coefficients in the effective theory of long wavelength modes, see eq. \eqref{eq: Wilson coefficients}.

We consider the Bunch-Davis vacuum, where inflationary modes are solved by the wave equation in the exact dS (we use the conformal time $\eta H_0=-e^{-H_0 t}$)
\begin{align}
\phi(\eta,x)=\int \frac{d^3k}{(2\pi)^3} \left(a_k u_k^-(\eta)e^{ik\cdot x}+a_k^\dagger u_k^+(\eta)e^{-ik\cdot x}\right)\,,\quad u_k^\pm(\eta)=\frac{H_0}{\sqrt{2k^3}}\left(1\mp ik\eta\right)e^{\pm ik\eta}\,,\label{eq: flat modes}
\end{align}
where $[a_k,a_{k'}^\dagger]=(2\pi)^{d-1}\delta^{d-1}(k-k')$.
The propagator in \eqref{eq: propagators} can thus be constructed by $\phi_e=\phi \,\theta(k+\varepsilon/\eta)$ and
\be
\langle \phi^c(t_1,x_1)\phi^c(t_2,x_2)\rangle=\frac{1}{2}\langle \phi(t_1,x_1)\phi(t_2,x_2)+\phi(t_2,x_2)\phi(t_1,x_1)\rangle\,,\quad \langle\phi^c(t_1,x_1)\phi^q(t_2,x_2)\rangle=\langle[\phi(t_1,x_1),\phi(t_2,x_2)]\rangle\theta(t_1-t_2)\,.
\ee
\subsection{From Feynman diagrams}

We start with the leading order diffusion, i.e., the first diagram in \eqref{eq: Feynman diagrams}, which is discussed extensively in the literature. 

This diagram gives
\be
\begin{tikzpicture}
[baseline={(0,-0.5ex)}]
\draw[thick,dashed] (-0.5,0.5)--(0,0);
\draw[thick] (0,0)--(1,0);
%\draw[thick,->] (0,0)--(0.6,0);
\draw[thick,dashed] (1,0)--(1.5,0.5);
\node at (-0.05,0.25) {$q$};
\node at (1+0.05,0.25) {$q$};
\node at (0.15,0.15) {$c$};
\node at (0.85,0.15) {$c$};
\node at (0,-0.3) {$t_1$};
\node at (1,-0.3) {$t_2$};
\end{tikzpicture}=\int dt_1 dt_2 a_1^3 a_2^3 \dot{\phi}_s^q(t_1,x_1)\dot{\phi}_s^q(t_2,x_2)\times \int \frac{d^3k}{(2\pi)^3}\partial_{t_1}\partial_{t_2} G^{cc}(t_1,t_2,k)\,,\label{eq: leading diff}
\ee
where $a_i=a(t_i)$ and $G^{cc}_{e}$ is the Keldysh propagator defined in \eqref{eq: propagators}. The term proportional to $k^2$ vanishes identically, because the system contracts directly with the environment in an orthogonal way: $\theta(k+\varepsilon/\eta)\,\theta(\varepsilon/\eta) \equiv 0$. In contrast, the contribution in \eqref{eq: leading diff} is non-zero, because the time derivative acts to produce a delta function $\delta(k+\varepsilon/\eta)$. This implies that short modes mix nontrivially with long modes near the “horizon” $k = -\varepsilon/\eta$, leading to diffusion. This is precisely why the standard Wilsonian EFT does not contain diffusion: there is no time-dependent scale to enforce UV–IR mixing. This diagram can be further decomposed into three contributions:
\begin{align}
& \int dt_1 dt_2 d^3x_1 d^3x_2 a_1^3 a_2^3 \dot{\phi}_s^q(t_1,x_1)\dot{\phi}_s^q(t_2,x_2) \times \int \frac{d^3k}{(2\pi)^3}\partial_{t_1}\partial_{t_2} G_e^{cc} (t_1,t_2,k)e^{ik\cdot x_{12}}\nonumber\\
& =\int dt_1 dt_2 d^3x_1 d^3x_2a_1^3 a_2^3 \dot{\phi}_s^q(t_1,x_1)\dot{\phi}_s^q(t_2,x_2)\times \int \frac{d^3k}{(2\pi)^3}\varepsilon^2 \dot{a}_1\dot{a}_2 H_0^2 \delta(k-a_1 H_0)\delta(k-a_2 H_0) G^{cc}(t_1,t_2,k)e^{ik\cdot x_{12}}\nonumber\\
&-2\int dt_1 dt_2d^3x_1 d^3x_2 a_1^3 a_2^3 \dot{\phi}^q_s(t_1,x_1)\phi^q_s(t_2,x_2)\times \int \frac{d^3k}{(2\pi)^3}\varepsilon^2 \dot{a}_1\dot{a}_2 H_0^2 \delta(k-a_1 H_0)\delta(k-a_2 H_0)\partial_{t_2}G^{cc}(t_1,t_2,k)e^{ik\cdot x_{12}}\nonumber\\
&+\int dt_1 dt_2 d^3x_1 d^3x_2 a_1^3 a_2^3 \phi^q_s(t_1,x_1)\phi^q_s(t_2,x_2)\times \int \frac{d^3k}{(2\pi)^3}\varepsilon^2 \dot{a}_1\dot{a}_2 H_0^2 \delta(k-a_1 H_0)\delta(k-a_2 H_0) \partial_{t_1}\partial_{t_2}G^{cc}(t_1,t_2,k)e^{ik\cdot x_{12}}\,,\label{eq: diffusion computation}
\end{align}
where $G^{cc}$ denotes the Keldysh propagator in \eqref{eq: propagators} without the projectors $\theta(k+\varepsilon/\eta)$. The second line and third line of \eqref{eq: diffusion computation} are identically zero, however, they support finite contributions in higher-order diagrams in \eqref{eq: Feynman diagrams}. As discussed extensively in the literature, we can easily compute the first line of \eqref{eq: diffusion computation} to give $H_0^3/(4\pi^2)$ at $\varepsilon\rightarrow 0$. This thus gives the influence functional with the leading diffusion effect
\be
S_{\rm IF}\supset i\frac{H_0^3}{8\pi^2}\int dt d^3x_1 d^3x_2 a^6  \dot{\phi}_s^q(t,x_1)\dot{\phi}_s^q(t,x_2)\,.
\ee
We note that this action is in fact non-local in space. Nevertheless, since $\phi^q_s$ has support only at extremely long wavelengths, the integral $\int d^3x\, \phi^q_s(t,x)$ effectively projects onto the zero mode, which we denote simply as $\phi^q_s(t)$, omitting the subscript $k = 0$ for convenience. In Starobinsky’s original formulation of stochastic inflation, this term is recast as $\xi(t) \phi^q_s(t)$, where $\xi(t)$ is a white noise \cite{Starobinsky:1986fx}. This corresponds to an annealed disorder, describing a stochastic force acting on the long-wavelength inflaton. We now show that this annealed disorder arises as an effective description, with a clear origin in the full theory. This influence functional mediates interactions between bra geometry and ket geometry, which may be related to the bra-ket wormhole \cite{Chen:2020tes,Fumagalli:2024msi}. We also note that upon Wick rotation to Euclidean AdS, the disorder becomes quenched, taking the form of Coleman-Giddings-Strominger-wormholes that generate bilocal operators \cite{coleman1988black,giddings1988axion,giddings1988loss,Chandra:2022bqq}.

We can now move to the second diagram in \eqref{eq: Feynman diagrams}. We can expand the vertices just as in \eqref{eq: diffusion computation}, and we find the term $(\phi^q_s)^2$ also gives a zero contribution in the leading multipole expansion $v''(\phi_s^c(t_3))\simeq v''(\phi_s^c(t_1))$. The non-vanishing contributions are (where we slip off $\int dx$ for simplicity by noting that in the end they extract the zero modes):
\begin{align}
& \begin{tikzpicture}[baseline={(0,-0.5ex)}]
\draw[thick,dashed] (-0.5,0.5)--(0,0);
\draw[thick] (0,0)--(0.65,0);
%\draw[thick,->] (0,0)--(0.45,0);
%\draw[thick,->] (0.85,0)--(1.25,0);
\draw[thick] (0.85,0)--(1.5,0);
\draw[thick] (0.75,0) circle (0.1);
   \draw[thick] (0.75-0.07, 0-0.07) -- (0.75+0.07, 0+0.07);
  \draw[thick] (0.75-0.07, 0+0.07) -- (0.75+0.07, 0-0.07);
\draw[thick,dashed] (1.5,0)--(2,0.5);
\draw[thick,dashed] (0.75,0.1) -- (0.75,0.5);
\node at (-0.05,0.25) {$q$};
\node at (1+0.05+0.5,0.25) {$q$};
\node at (0.15,0.15) {$c$};
\node at (0.85+0.5,0.15) {$c$};
\node at (0,-0.3) {$t_1$};
\node at (1+0.5,-0.3) {$t_2$};
\node at (0.75,0.6) {$c$};
\node at (0.75,-0.3) {$t_3$};
\node at (0.55,0.15) {$c$};
\node at (0.95,0.15) {$q$};
\end{tikzpicture}=\nonumber\\
& \int dt_1dt_2dt_3a_1^3 a_2^3 a_3^3 v''(\phi_s^c(t_1))\dot{\phi}_s^q(t_1)\dot{\phi}_s^q(t_2)\times \int \frac{d^3k}{(2\pi)^3}\varepsilon^2 \dot{a}_1\dot{a}_2 H_0^2 \delta(k-a_1 H_0)\delta(k-a_2 H_0) G^{cc}(t_1,t_3,k)G^{cq}(t_2,t_3,k)e^{ik\cdot x_{12}}\,,\nonumber\\
& 2\int dt_1dt_2dt_3a_1^3 a_2^3 a_3^3 v''(\phi_s^c(t_1))\dot{\phi}_s^q(t_1)\dot{\phi}_s^q(t_2)\times \int \frac{d^3k}{(2\pi)^3}\varepsilon^2 \dot{a}_1\dot{a}_2 H_0^2 \delta(k-a_1 H_0)\delta(k-a_2 H_0) G^{cc}(t_1,t_3,k)\partial_{t_2}G^{cq}(t_2,t_3,k)e^{ik\cdot x_{12}}\,,
\end{align}
where we use the multipole expansion in $x$ to set $k_2=k_1=k$ and move $v''(\phi^c_s(t,x_3))\simeq v''(\phi^c_s(t,x_1))$. Computing the integral gives the influence functional for zero modes
\be
S_{\rm IF}\supset i \frac{H_0}{12\pi^2}\int dt a^6 v''(\phi_s^c(t)) \left( \dot{\phi}^q_s(t)\dot{\phi}^q_s(t)\log+H_0\dot{\phi}^q_s(t)\phi^q_s(t)\right)\,,
\ee
where $\log=\log\frac{\varepsilon}{2}-\psi\left(\frac{3}{2}\right)$.

We now compute The third and fourth diagrams in \eqref{eq: Feynman diagrams}. It is useful to define
\begin{align}
F(t_1,t_2)&=2\int dt_3 dt_4 a_1^3 a_2^3 a_3^3 a_4^3 \times  \int \frac{d^3k}{(2\pi)^3}\varepsilon^2 \dot{a}_1\dot{a_2} H_0^2 \delta(k-a_1 H_0)\delta(k-a_2 H_0)G^{cc}(t_1,t_3,k)G^{cq}(t_4,t_3,k)G^{cq}(t_2,t_4,k)\nonumber\\
&+\int dt_3 dt_4 a_1^3 a_2^3 a_3^3 a_4^3 \times  \int \frac{d^3k}{(2\pi)^3}\varepsilon^2 \dot{a}_1\dot{a_2} H_0^2 \delta(k-a_1 H_0)\delta(k-a_2 H_0)G^{cq}(t_1,t_3,k)G^{cc}(t_3,t_4,k)G^{cq}(t_2,t_4,k)\,.
\end{align}
We then have
\begin{align}
2\times
&\begin{tikzpicture}[baseline={(0,-0.5ex)}]
%\draw[thick,->] (0,0)--(0.45,0);
%\draw[thick,->] (0.85,0)--(1.25,0);
%\draw[thick,->] (1.6,0) --(2,0);
\draw[thick,dashed] (-0.5,0.5)--(0,0);
\draw[thick] (0,0)--(0.65,0);
\draw[thick] (0.85,0)--(1.4,0);
\draw[thick] (1.6,0)--(2.25,0);
\draw[thick] (0.75,0) circle (0.1);
   \draw[thick] (0.75-0.07, 0-0.07) -- (0.75+0.07, 0+0.07);
  \draw[thick] (0.75-0.07, 0+0.07) -- (0.75+0.07, 0-0.07);
\draw[thick,dashed] (2.25,0)--(2.75,0.5);
\draw[thick,dashed] (0.75,0.1) -- (0.75,0.5);
\draw[thick] (1.5,0) circle (0.1);
   \draw[thick] (1.5-0.07, 0-0.07) -- (1.5+0.07, 0+0.07);
  \draw[thick] (1.5-0.07, 0+0.07) -- (1.5+0.07, 0-0.07);
\draw[thick,dashed] (2.25,0)--(2.75,0.5);
\draw[thick,dashed] (0.75,0.1) -- (0.75,0.5);
\draw[thick,dashed] (1.5,0.1) -- (1.5,0.5);
\node at (-0.05,0.25) {$q$};
\node at (1+0.05+0.5+0.75,0.25) {$q$};
\node at (0.15,0.15) {$c$};
\node at (0.85+0.5+0.75,0.15) {$c$};
\node at (0,-0.3) {$t_1$};
\node at (1+0.5+0.75,-0.3) {$t_2$};
\node at (0.75,0.6) {$c$};
\node at (1.5,0.6) {$c$};
\node at (0.75,-0.3) {$t_3$};
\node at (1.5,-0.3) {$t_4$};
\node at (0.55,0.15) {$c$};
\node at (0.95,0.15) {$q$};
\node at (1.3,0.15) {$c$};
\node at (1.7,0.15) {$q$};
\end{tikzpicture}+
\begin{tikzpicture}[baseline={(0,-0.5ex)}]
%\draw[thick,->] (0,0)--(0.45,0);
%\draw[thick,->] (0.85,0)--(1.25,0);
%\draw[thick,->] (1.6,0) --(2,0);
\draw[thick,dashed] (-0.5,0.5)--(0,0);
\draw[thick] (0,0)--(0.65,0);
\draw[thick] (0.85,0)--(1.4,0);
\draw[thick] (1.6,0)--(2.25,0);
\draw[thick] (0.75,0) circle (0.1);
   \draw[thick] (0.75-0.07, 0-0.07) -- (0.75+0.07, 0+0.07);
  \draw[thick] (0.75-0.07, 0+0.07) -- (0.75+0.07, 0-0.07);
\draw[thick,dashed] (2.25,0)--(2.75,0.5);
\draw[thick,dashed] (0.75,0.1) -- (0.75,0.5);
\draw[thick] (1.5,0) circle (0.1);
   \draw[thick] (1.5-0.07, 0-0.07) -- (1.5+0.07, 0+0.07);
  \draw[thick] (1.5-0.07, 0+0.07) -- (1.5+0.07, 0-0.07);
\draw[thick,dashed] (2.25,0)--(2.75,0.5);
\draw[thick,dashed] (0.75,0.1) -- (0.75,0.5);
\draw[thick,dashed] (1.5,0.1) -- (1.5,0.5);
\node at (-0.05,0.25) {$q$};
\node at (1+0.05+0.5+0.75,0.25) {$q$};
\node at (0.15,0.15) {$c$};
\node at (0.85+0.5+0.75,0.15) {$c$};
\node at (0,-0.3) {$t_1$};
\node at (1+0.5+0.75,-0.3) {$t_2$};
\node at (0.75,0.6) {$c$};
\node at (1.5,0.6) {$c$};
\node at (0.75,-0.3) {$t_3$};
\node at (1.5,-0.3) {$t_4$};
\node at (0.55,0.15) {$q$};
\node at (0.95,0.15) {$c$};
\node at (1.3,0.15) {$c$};
\node at (1.7,0.15) {$q$};
\end{tikzpicture}\nonumber\\
&=\int dt_1dt_2 v''(\phi^c_s(t_1))^2\left( \dot{\phi}^q_s(t_1)\dot{\phi}^q_s(t_2)-2\dot{\phi}^q_s(t_1)\dot{\phi}^q_s(t_2)\partial_{t_2}+\phi^q_s(t_1)\phi^q_s(t_2)\partial_{t_1}\partial_{t_2}\right)F(t_1,t_2)\,.
\end{align}
This leads to
\be
S_{\rm IF}\supset i \frac{1}{36 H_0 \pi^2} \int dt\, a^6 v''(\phi_s^c)^2\left(\left(\log^2+\frac{\log}{3}+\frac{\pi^2-8}{4}\right)(\dot{\phi}^q_s)^2+2H_0\left(\log+\frac{1}{6}\right)\dot{\phi}^q_s\phi^q_s+\frac{H_0^2}{2} (\phi^q_s)^2\right)\,.
\ee

Note that there is also a vertex such as $\begin{tikzpicture}[baseline={(0,-0.5ex)}]
\draw[thick,dashed] (0,0) -- (0.7,0);
  \draw[thick] (0.7,0) -- (1.4,0);
  \node at (0.6,0.2) {$c$};
  \node at (0.8,0.2) {$q$};
  \node at (0.7,-0.3) {$t$};
\end{tikzpicture}$, thus we should also compute relevant diagrams by replacing the vertex, for example $\begin{tikzpicture}[baseline={(0,-0.5ex)}]
%\draw[thick,->] (0,0)--(0.45,0);
%\draw[thick,->] (0.85,0)--(1.25,0);
%\draw[thick,->] (1.6,0) --(2,0);
\draw[thick,dashed] (-0.5,0.5)--(0,0);
\draw[thick] (0,0)--(0.65,0);
\draw[thick] (0.85,0)--(1.4,0);
\draw[thick] (1.6,0)--(2.25,0);
\draw[thick] (0.75,0) circle (0.1);
   \draw[thick] (0.75-0.07, 0-0.07) -- (0.75+0.07, 0+0.07);
  \draw[thick] (0.75-0.07, 0+0.07) -- (0.75+0.07, 0-0.07);
\draw[thick,dashed] (2.25,0)--(2.75,0.5);
\draw[thick,dashed] (0.75,0.1) -- (0.75,0.5);
\draw[thick] (1.5,0) circle (0.1);
   \draw[thick] (1.5-0.07, 0-0.07) -- (1.5+0.07, 0+0.07);
  \draw[thick] (1.5-0.07, 0+0.07) -- (1.5+0.07, 0-0.07);
\draw[thick,dashed] (2.25,0)--(2.75,0.5);
\draw[thick,dashed] (0.75,0.1) -- (0.75,0.5);
\draw[thick,dashed] (1.5,0.1) -- (1.5,0.5);
\node at (-0.05,0.25) {$c$};
\node at (1+0.05+0.5+0.75,0.25) {$q$};
\node at (0.15,0.15) {$q$};
\node at (0.85+0.5+0.75,0.15) {$c$};
\node at (0,-0.3) {$t_1$};
\node at (1+0.5+0.75,-0.3) {$t_2$};
\node at (0.75,0.6) {$c$};
\node at (1.5,0.6) {$c$};
\node at (0.75,-0.3) {$t_3$};
\node at (1.5,-0.3) {$t_4$};
\node at (0.55,0.15) {$c$};
\node at (0.95,0.15) {$q$};
\node at (1.3,0.15) {$c$};
\node at (1.7,0.15) {$q$};
\end{tikzpicture}$. Nevertheless, diagrams  with this vertex are identically zero in Markovian approximation, because these diagrams are all time-ordered, such as $t_2>t_4>t_3>t_1$, but the Markovian scale separation sets $t_1=t_2$. See also \cite{Tokuda:2017fdh} for an argument that there is indeed no such response coefficients.

This completes the derivation of \eqref{eq: Wilson coefficients}. 

\subsection{From massive correlators}

An alternative way is to consider the time multipole expansion from the beginning, treating $v''(\phi^c_s)$ as a static source serving as a mass term for short modes. This approach was adopted in \cite{gorbenko2019lambda}. Then, the diffusion terms can be derived from \eqref{eq: diffusion computation} but inserting massive propagators with $m_s^2=v''$. This is also known as the background field method in EFTs.

With a finite mass $m_s$, the inflationary modes become:
\be
u_k^\pm(\eta)=%\pmi
-\frac{H_0\sqrt{\pi} }{2}(-\eta)^{\frac{3}{2}}e^{\mp i \pi \nu_{s}}H^{\mp}_{-\nu_s}\left(k\eta\right)\,,
\ee
where $H^{+}_{\nu_s}=H^{(1)}_{\nu_s}$ is the first kind Hankel function and $H^{-}_{\nu_s}=H^{(2)}_{\nu_s}$ is the second kind, with $\nu_s=3/2\sqrt{1-(2m_s/(3H_0))^2}$. The propgators are given by:
\begin{align}
& G^{cc}(\eta_1,\eta_2,k)=\frac{ \pi}{4}  \eta _1^{3/2} \eta _2^{3/2} H_0^2 \left(J_{\nu _s}\left(k \eta _1\right) J_{\nu _s}\left(k \eta _2\right)+Y_{\nu _s}\left(k \eta _1\right) Y_{\nu _s}\left(k \eta _2\right)\right)\,,\nonumber\\
&G^{cq}(\eta_1,\eta_2,k)=-\frac{i \pi}{2}   \eta _1^{3/2} \eta _2^{3/2} H_0^2 \left(J_{\nu _s}\left(k \eta _1\right) J_{\nu _s}\left(k \eta _2\right)-Y_{\nu _s}\left(k \eta _1\right) Y_{\nu _s}\left(k \eta _2\right)\right)\theta(t_1-t_2)\,.
\end{align}
Substituting this into \eqref{eq: diffusion computation}, we find \eqref{eq: EFT} with
\begin{align}
& C_1(\phi^c_s)=\frac{H_0^3 4^{\nu_s(\phi^c_s)-2} \varepsilon ^{3-2\nu_s(\phi^c_s)} \Gamma \left(\nu_s(\phi^c_s)\right)^2}{\pi ^3}\,,\quad C_2(\phi_s^c)=\frac{H_0^4 4^{\nu _s\left(\phi _s^c\right)-2} \left(2 \nu _s\left(\phi _s^c\right)-3\right) \varepsilon ^{3-2 \nu _s\left(\phi _s^c\right)} \Gamma \left(\nu _s\left(\phi _s^c\right)\right){}^2}{\pi ^3}\,,\nonumber\\
& C_2(\phi^c_s)=\frac{H_0^5 4^{\nu _s\left(\phi _s^c\right)-3} \left(3-2 \nu _s\left(\phi _s^c\right)\right){}^2 \varepsilon ^{3-2 \nu _s\left(\phi _s^c\right)} \Gamma \left(\nu _s\left(\phi _s^c\right)\right){}^2}{\pi ^3}\,.\label{eq: coefficients exact}
\end{align}
Expanding these coefficients around $m_s\rightarrow 0$ to $v''(\phi^c_s)^2$ order, we find the diffusion coefficients  \eqref{eq: Wilson coefficients}. We can think of \eqref{eq: coefficients exact} as improved diffusion coefficients, as they resum subset of diagrams of the type in \eqref{eq: Feynman diagrams} to all orders in $v''(\phi^c_s)$. We can then also claim improved Fokker-Planck equation \eqref{eq: FP} and its phase space generalization \eqref{eq: Klein-Kramers} using \eqref{eq: coefficients exact}. %which can be achieved by renormalizing the IR divergence $\log\varepsilon|_{\varepsilon\rightarrow 0}$ with an anomalous dimension $3-2\nu_s(\phi^c_s)$.

\section{Effects from deviating an exact dS}

In this material, we present the details of incorporating the effects of a background that deviates from exact dS space. We only consider the leading order deviation.

To consider a dynamical $a^c$, which deviates from $a_0^c=e^{H_0 t}$ by back-reactions, we assume it can be determined by trajectory of long wavelength modes as  we trace out the environment. 
\be
a=e^{H_0 t}\left(1+\fft{1}{6H_0^2}\delta_a(\phi^c_s)\right)\,,\quad \delta_a(\phi^c_s)=\int^t dt' \delta(\phi^c_s(t'))\,,\label{eq: expand a}
\ee
where we take $\delta,\delta_a\ll 1$.

We now show that this deviation provides new vertex for tracing out the short modes. 
It is important to note that we should not directly expand the action using \eqref{eq: expand a}, otherwise the information of dynamical behavior of short modes will be lost. The strategy is to consider
\be
-(\nabla\phi)^2\rightarrow \phi\Box \phi\,,
\ee
followed by expanding \eqref{eq: expand a} and integrating by parts back the standard form of the action. This prescription keeps the dynamical information of short modes intact upon expanding $a^c$.

The Liouvelle action for environment perturbatively expanded in $\delta$ is
\begin{align}
S_{L,{\phi_e}}&=\int dt d^3x\left((a^c_0)^3 \dot{\phi}^c_e \dot{\phi}^q_e-a^c_0 \partial_i \phi^c_e \partial^i \phi^q_e\right)\nonumber\\
& -\fft{1}{12 H_0^2} \int dtd^3x \left(-6\delta_a(\phi_s^c) (a^c_0)^3 \dot{\phi}^c_e \dot{\phi}^q_e +2 \delta_a(\phi_s^c)a^c_0 \partial_i \phi^c_e \partial^i \phi^q_e-9\delta(\phi_s^c) (a^c_0)^3 H_0^2 \phi^c_e \phi^q_e \right)\,.\label{eq: new action e}
\end{align}
This gives an additional ``mass'' term originated from the deviation from dS, i.e., $-3/8 (a^c_0)^3 \delta(\phi^c_s) \phi^c_e\phi^q_e$. It is not necessary to expand the interaction terms between $\phi_e$ and $\phi_s$ in $\delta, \delta_a$. However, the action \eqref{eq: new action e} possesses an issue for perturbation theory, because the interaction term modifies kinetic term by $\dot{\phi}^c_e \dot{\phi}^q_e$. To deal with this, we can make a field redefinition
\be
\phi_e\rightarrow \fft{\phi_e}{\sqrt{\fft{2H_0^2+\delta_a(\phi^c_s)}{2H_0^2}}}\,.
\ee
The resulting action is:
\begin{align}
& S_{L,{\phi_e}}=\int dt d^3x\left((a_0)^3 \dot{\phi}^c_e \dot{\phi}^q_e-a_0 \partial_i \phi^c_e \partial^i \phi^q_e\right)-\fft{1}{12 H_0^2} \int dtd^3x \left(4\delta_a(\phi_s^c) a_0 \partial_i \phi^c_e \partial^i \phi^q_e-9 \delta(\phi_s^c) (a_0)^3 H_0^2 \phi^c_e \phi^q_e \right)\,,\nonumber\\
& S_{L,{\rm int}}=\int dt d^3x  \left(a^3\dot{\phi}^c_{(e} \dot{\phi}^q_{s)}-a\partial_i\phi^c_{(e} \partial^i\phi^q_{s)}\right)\left(1- \fft{\delta_a(\phi^c_s)}{4H_0^2}\right)+\cdots\,,\label{eq: shift coupling}
\end{align}
where we only keep the relevant term for including the deviation effect. This action modifies an existing vertex and introduces a new vertex (where we drop $k^2$ term for simplicity as it contributes nothing eventually)
\begin{align}
& \begin{tikzpicture}[baseline={(0,-0.5ex)}]
\draw[thick,dashed] (0,0) -- (0.7,0);
  \draw[thick] (0.7,0) -- (1.4,0);
  \node at (0.6,0.2) {$q$};
  \node at (0.8,0.2) {$c$};
  \node at (0.7,-0.3) {$t$};
\end{tikzpicture}=\left(1-\frac{\delta_a(\phi_s^c)}{4H_0^2}\right)\left(a^3 \dot{\phi}_s^q \partial_t-a k^2 \phi_s^q\right)\,,\quad
\begin{tikzpicture}[baseline={(0,-0.5ex)}]
  \draw[thick] (0,0) -- (0.6,0);
   \draw[thick] (0.8,0) -- (1.4,0);
  \fill (0.7,0) circle (0.1);
   \draw[thick] (0.7-0.07, 0-0.07) -- (0.7+0.07, 0+0.07);
  \draw[thick] (0.7-0.07, 0+0.07) -- (0.7+0.07, 0-0.07);
  \node at (0.5,0.2) {$c$};
  \node at (0.9,0.2) {$q$};
  \draw[thick,dashed] (0.7,0.1) -- (0.7,0.6);
  \node at (0.7,0.7) {$c$};
  \node at (0.7,-0.3) {$t$};
\end{tikzpicture}= \fft{a_0}{12H_0^2}\left(9a_0^2 \delta(\phi_c^s)-4 k^2 \delta_a(\phi_c^s)\right)\,.
\end{align}
Other vertices are also modified by the field redefinition, however, these effects appear at higher-order as $\mathcal{O}(v \delta)$ that we do not consider. We leave more complete corrections to future work. 

We can compute the new Feynman diagram and then find
\be
\begin{tikzpicture}
[baseline={(0,-0.5ex)}]
\draw[thick,dashed] (-0.5,0.5)--(0,0);
\draw[thick] (0,0)--(1,0);
\draw[thick,->] (0,0)--(0.6,0);
\draw[thick,dashed] (1,0)--(1.5,0.5);
\node at (-0.05,0.25) {$q$};
\node at (1+0.05,0.25) {$q$};
\node at (0.15,0.15) {$c$};
\node at (0.85,0.15) {$c$};
\node at (0,-0.3) {$t_1$};
\node at (1,-0.3) {$t_2$};
\end{tikzpicture}+2\times \begin{tikzpicture}[baseline={(0,-0.5ex)}]
\draw[thick,dashed] (-0.5,0.5)--(0,0);
\draw[thick] (0,0)--(0.65,0);
\draw[thick,->] (0,0)--(0.45,0);
\draw[thick,->] (0.85,0)--(1.25,0);
\draw[thick] (0.85,0)--(1.5,0);
\fill (0.75,0) circle (0.1);
   \draw[thick] (0.75-0.07, 0-0.07) -- (0.75+0.07, 0+0.07);
  \draw[thick] (0.75-0.07, 0+0.07) -- (0.75+0.07, 0-0.07);
\draw[thick,dashed] (1.5,0)--(2,0.5);
\draw[thick,dashed] (0.75,0.1) -- (0.75,0.5);
\node at (-0.05,0.25) {$q$};
\node at (1+0.05+0.5,0.25) {$q$};
\node at (0.15,0.15) {$c$};
\node at (0.85+0.5,0.15) {$c$};
\node at (0,-0.3) {$t_1$};
\node at (1+0.5,-0.3) {$t_2$};
\node at (0.75,0.6) {$c$};
\node at (0.75,-0.3) {$t_3$};
\node at (0.55,0.15) {$c$};
\node at (0.95,0.15) {$q$};
\end{tikzpicture}\rightarrow S_{\rm IF}\supset -i \frac{H_0}{16\pi^2}\int dt a^6 \delta(\phi_s^c(t)) \left( \dot{\phi}^q_s(t)\dot{\phi}^q_s(t)\log+H_0\dot{\phi}^q_s(t)\phi^q_s(t)\right)\,.
\ee
Thus, we simply find a correction to $C_1,C_2$ by shifting $v''\rightarrow v''-\frac{3}{4}\delta$ and retaining only terms below the order $\mathcal{O}(v \delta)$.

However, we should emphasize that although $\delta(\phi^c_s)$ enters the diffusion coefficients, it is unknown for now. Nevertheless, it can be determined by solving the inflationary trajectory in effective theory $S_L^{\rm eff}$. The equation of motion can be obtain by the variation principle. Noting $S_L^{\rm eff}=\int dt \mathcal{L}_L^{\rm eff}$, we need to solve
\be
\frac{\delta \mathcal{L}_L^{\rm eff}}{\delta N^q}=0\,,\quad \frac{\delta \mathcal{L}_L^{\rm eff}}{\delta \phi^{q/c}}=0\,,\quad \frac{\delta \mathcal{L}_L^{\rm eff}}{\delta a^{q/c}}=0\,,
\ee
at the time surface where we evaluate the density matrix, subjecting to any boundary condition we choose to project the density matrix. In semi-classical gravitational regime, we always consider the boundary condition $a^q=N^q=0$. These equations are:
\begin{align}
& \frac{1}{2}\left(\dot{\phi}^c\right)^2+V_r(\phi^c)-3H^2=0\,,\quad \frac{1}{2} a^2 (\dot{\phi}^c)^2+2 a\, \ddot{a}+\dot{a}^2-a^2 V_r(\phi ^c)=0\,,\nonumber\\
&\ddot{\phi}^c+3H\dot{\phi}^c+ v_r'(\phi^c)+ i a(t)^3 \left(2C_1(\phi^c)\left(\ddot{\phi}^q+6H\dot{\phi}^q\right)+\phi^q\left(6HC_2(\phi^c)-2C_3(\phi^c)+C_2'(\phi^c)\dot{\phi}^c\right)+2C_1'(\phi^c)\dot{\phi}^c\dot{\phi^q}\right)=0\,,\nonumber\\
& \dot{\phi}^c \dot{\phi}^q-\dot{\phi}^q V_r'(\phi^c)+2ia^3\left(C_1(\phi^c)\dot{\phi}^q+C_2(\phi^c)\dot{\phi}^q\phi^q+C_3(\phi^c)(\phi^q)^2\right)=0\,,\nonumber\\
& \ddot{\phi}^q+3H\dot{\phi}^q+ v_r''(\phi^c)\phi^q-ia^3\left(C_1'(\phi^c)(\dot{\phi}^q)^2+C_2'(\phi^c)\dot{\phi}^q\phi^q+C_3'(\phi^c)(\phi^q)^2\right)=0\,,
\end{align}
where $H=\dot{a}/a$.

In general, solving these equations is difficult, but we only need the solution at the time when the density matrix is measured, given the specified boundary conditions. 

For deriving Fokker-Planck equation, we set $\phi^q(t)=0$ (here $t$ is the time where we evaluate the density matrix rather than the history). Then slow-roll approximation for $\phi^q$ also sets $\dot{\phi}^q=0$, which simplifies the problem to be:
\be
\frac{1}{2}\left(\dot{\phi}^c\right)^2+V_r(\phi^c)-3H^2=0\,,\quad \ddot{\phi}^c+3H\dot{\phi}^c+ v_r'(\phi^c)=0\,.
\ee
To solve these equations, we scale 
\be
v_r=\kappa v_r \,,\quad \partial_t^i\phi^c=\sum_{n=0}\kappa^{n+i}(\partial_t^i \phi^c)_n\,,\quad  H=H_0+\sum_{n=1} \kappa^n H_n\,,\quad a=a_0+\sum_{n=1}\kappa^n a_n\,,
\ee
with $\kappa\rightarrow 0$ and fixed $\phi^c$.
We can easily find the solution in the slow-roll limit
\begin{align}
&H(\phi^c)=H_0+\fft{v_r(\phi^c)}{6H_0}+\fft{2V_r'(\phi^c)^2-3v_r(\phi^c)^2}{216H_0^3}\cdots\,,\quad
\dot{\phi}^c=-\fft{V_r'(\phi^c)}{3H_0}+\fft{v_r(\phi^c)V_r'(\phi^c)}{18 H_0^3}-\fft{V_r'(\phi^c)V_r''(\phi^c)}{27 H_0^3}+\cdots\,,\nonumber\\
%\delta(\phi^c)=V_r(\phi^c)-V_0=v_r(\phi^c)\,,\nonumber\\
& a=a_0\left(1+\fft{\int^t  dt'v_r(\phi^c(t'))}{6H_0^2}+\fft{2\int^t dt' v_r'(\phi^c(t'))^2 -3\int^t dt' v_r(\phi^c(t'))^2+6\int^t dt' \int^{t'} dt''v_r(\phi^c(t')v_r(\phi^c(t''))}{216 H_0^3}+\cdots\right)\,.
\end{align}

Determining $\delta(\phi^c_s)$ in Wigner function is slightly more nontrivial, nevertheless we find the same solution $\delta=v_r$ at leading order, while the next-to-leading order is significantly different. We consider the scaling:
\begin{align}
& v_r=\kappa v_r \,,\quad \partial_t^i\phi^c=\sum_{n=0}\kappa^{n+i}(\partial_t^i \phi^c)_n\,,\quad  H=H_0+\sum_{n=1} \kappa^n H_n\,,\quad a=a_0+\sum_{n=1}\kappa^n a_n\,,\nonumber\\
& \phi^q=\sum_{n=1}\kappa^n \phi^q_n\,,\quad \partial_t^i \phi^q=\sum_{n=1} \kappa^{n+i}\partial_t^i \phi^q_n\,,\quad \Pi^c=\kappa \Pi^c_0+\kappa^2 \Pi^c_1\,,\quad \partial_t^i \Pi^c=\sum_{n=0} \kappa^{n+i} (\partial_t^i \Pi^c)_n\,,
\end{align}
where we now fix $\phi^c,\Pi_1^c$. Note that $\Pi_0^c$ is determined by the equations because, in the semi-classical limit, the density matrix is concentrated near the momentum equilibrium. We find, for example,
\be
\Pi_0^c=-a_0^3 v_r'(\phi^c)\,,\quad H(\phi^c)=H_0+\fft{v_r(\phi^c)}{6H_0}+\fft{6H_0^2 (\dot{\phi}^c)^2-v_r(\phi^c)^2}{72H_0^3}+\cdots\,,\quad \dot{\phi}^c=-\fft{v_r'(\phi^c)}{3H_0}+6i a_0^3 H_0 C_1(\phi^c)\phi^q+\cdots\,.
\ee
We didn't present the solutions of $\phi^q, a-a_0, \partial_t \Pi^c$, as there are no simple closed forms. Nevertheless, we verified that the equations are consistent.

\section{Play with master equations}

\subsection{Equilibrium states of Fokker-Planck equation}

In this material, we solve the equilibrium state of FP equation in \eqref{eq: FP}. For simplicity, we ignore the renormalization effects $v_r=v$.

The equilibrium state is a steady state $\partial_t P$ without any flow of the probability current $J$. Namely, we can write the FP equation as $\partial_t P=-\partial_\phi J$ and require $J\equiv 0$. The leading order solution was shown in \cite{Starobinsky:1986fx,Starobinsky:1994bd}
\be
P_{\rm eq}(\phi)=\mathcal{N}e^{-\fft{8\pi^2 v(\phi)}{3H_0^4}}\,,
\ee
where $\mathcal{N}$ is the normalization to ensure $\int d\phi P_{\rm eq}(\phi)\equiv 1$.
This is identical to the squared of the Hartle-Hawking wavefunction \cite{hartle1983wave} by semi-classically expanding $v\ll H_0^2$
\be
P_{\rm HH}=|\Psi_{\rm HH}|^2\sim e^{\fft{24\pi^2}{V(\phi)}}\sim P_{{\rm HH},0} P_{\rm eq}(\phi)\,,
\ee
where $P_{{\rm HH},0}\sim e^{8\pi^2/H_0^2}$ is the squared Hartle-Hawking wavefunction at initial time, which can be computed by Euclidean path integral for a pure dS.

For \eqref{eq: FP} in the limit $v/H_0^2\ll1$, we find:
\begin{align}
& P_{\rm eq}(\phi)=\mathcal{N}e^{-\fft{8\pi^2 v(\phi)}{3H_0^4}+A_1(\phi)+A_2(\phi)}\,,\quad A_1(\phi)=\frac{  3 v(\phi )-4 v''(\phi )}{6 H_0^2}\log\,,\nonumber\\
& A_2(\phi)=\frac{-72 H_0^2 \log ^2 v(\phi ) v''(\phi )-4 H_0^2 \left(3 \pi ^2-24+4 \log \right) v''(\phi )^2+3 \left(9 H_0^2 \log ^2+16 \pi ^2 (1-3 \log )\right) v(\phi )^2+32 \pi ^2 (6 \log -1) v'(\phi )^2}{216 H_0^6}\,.
\end{align}

To determine $\mathcal{N}$, we need details of the potential $v(\phi)$. Then, we can evaluate the equal-time correlators. In this Letter, we only present one-point functions $\langle\phi^2\rangle$
\be
\langle \phi^2\rangle=\int d\phi \, \phi^2 P_{\rm eq}(\phi).
\ee
We consider polynomial potentials:
\begin{align}
& v=c |\phi|:\,\quad \langle\phi^2\rangle\simeq\fft{0.003 H_0}{c^2}\,,\nonumber\\
& v= \fft{m^2}{2}\phi^2\,:\quad \langle \phi^2\rangle\simeq\fft{0.038 H_0^4}{m^2}+H_0^2\left(0.025\log-0.004\right)\,,\nonumber\\
& v=\fft{g}{3} |\phi|^3\,:\quad \langle\phi^2\rangle\simeq \frac{0.088 H_0^{8/3}}{g^{2/3}}+H_0^2 (0.014 \log -0.006)+g^{2/3} H_0^{4/3} \left(-0.005+0.012 \log ^2-0.004 \log \right)\,,\nonumber\\
& v=\fft{\lambda}{4}\phi^4\,:\quad \langle\phi^2\rangle\simeq \frac{0.132 H_0^2}{\sqrt{\lambda }}+H_0^2 (0.005 \log -0.008)+H_0^2 \sqrt{\lambda } \left(-0.01+0.021 \log ^2-0.008 \log \right)\,,\nonumber\\
& v=\fft{g_5}{5}|\phi|^5\,:\quad \langle\phi^2\rangle\simeq \frac{0.167 H_0^{8/5}}{g_5^{2/5}}+H_0^2 (-0.009-0.004 \log )+g_5^{2/5} H_0^{12/5} \left(-0.017+0.034 \log ^2-0.013 \log \right)\,,\nonumber\\
& v=\fft{g_6}{6}\phi^6\,:\quad \langle\phi^2\rangle\simeq \frac{0.195 H_0^{4/3}}{g_6^{\fft{1}{3}}}+H_0^2 (-0.011-0.013 \log )+g_6^{\fft{1}{3}} H_0^{8/3} \left(-0.026+0.054 \log ^2-0.02 \log \right)\,.
\end{align}
Generally, we thus expect the scaling behavior for polynomial potential $c\sim g_n \phi^n$ as:
\be
\langle\phi^2\rangle\simeq \#_1 \fft{H_0^{\fft{8}{n}}}{g_n^{\fft{2}{n}}}+\#_2 H_0^2+ \#_3 H_0^{\fft{2(2n-4)}{n}}g_n^{\fft{2}{n}}+\cdots\,.
\ee

We leave the explorations of correlators $\langle\phi(t_1)\phi(t_2)\rangle$ as well as out-of-equilibrium states to future studies.

\subsection{From master equation of the Wigner function to Fokker-Planck equation}

In this material, we derive the FP equation \eqref{eq: FP} from the master equation in phase-space \eqref{eq: Klein-Kramers}. This is also a consistency check of our result. For simplicity, we consider exact dS.

The diagonal density matrix $P(\phi,t)$ can be obtained from the Wigner function by integrating out the conjugate momentum
\be
P(\phi,t)=\int d\Pi\, W(\phi,\Pi,t)\,.
\ee
In general, we can define $n$-th moment of the Wigner function
\be
W_n(\phi,t)=\int d\Pi \left(\fft{\Pi}{a^3}\right)^n W(\phi,\Pi,t)\,.
\ee
We now take the time derivative on both sides and then use the equation \eqref{eq: Klein-Kramers}. We find:
\be
\partial_t P(\phi,t) = -\partial_\phi W_1(\phi,,t)+\partial_\phi^2\left(C_1(\phi) P(\phi,t)\right)\,.
\ee
This is already in the form of \eqref{eq: FP}. Our goal is then to show that we indeed have $W_1=a^{-3} \Pi^c_{\rm sad}P$ in \eqref{eq: saddle momentum} in appropriate limits.

For this purpose, we consider recursive equations for all $n$-moments:
\be
\partial_t W_n+\partial_\phi W_{n+1}+n v_r' W_{n-1}+3 H_0 n W_n-n(n-1)C_3 W_{n-2}-\partial_\phi\left(\partial_\phi\left(C_1 W_n \right)+n C_2 W_{n-1}\right)=0\,.
\ee
To solve these equations, we consider the slow-roll limit in eternal inflation regime $H_0\sim \sqrt{\epsilon}$. It is important to note that the slow-roll limit plays a role of ``overdamping'' limit in Brownian motion. We take all slow-roll parameters as the same order in the first slow-roll parameter
\be
\epsilon^{(n)}\sim v^{(n)}/v\,,\quad\epsilon^{(n)}\sim\epsilon\,.
\ee
Then, a consistent power counting is roughly:
\be
v'\sim v_r'\sim \epsilon^{3/2}\,,\quad v^{(n)}\sim v_r^{(n)}\sim \epsilon^2\,,\quad W_n\sim \sum_{j=0} W_{nj}\epsilon^{j+n}\,,\quad \partial_\phi^n\sim \sum_{j=0}(\partial_\phi^n)_j \epsilon^{\fft{n+j}{2}}\,.
\ee
Accordingly, $\partial_t$ also scales in $\epsilon$, and its action on higher-$n$ moments is pushed to be in very high order; therefore, the higher Wigner moments are almost stationary. This power counting also pushes the effects of $C_{2,3}$ far away. 

We consider the order to $\epsilon^{5/2}$ and find the equations becoming:
\be
W_{20}+\fft{v_r'}{3H_0}W_{10}=0\,,\quad (\partial_\phi W_{20})_0+3H_0 W_{11}=0\,,\quad W_{10}+\fft{v_r'}{3H_0}P=0\,.
\ee
It is then straightforwardly to find
\be
W_1=W_{10}+W_{11}=-\left(\fft{v_r'}{3H_0}+\fft{v_r' v_r''}{(3H_0)^3}\right)P\,,
\ee
which is precisely $a^{-3}\Pi^c_{\rm sad}P$.

\section{Details in global slicing}

In this material, we provide more details of stochastic inflation in global slicing and derive its Fokker-Planck equation.

The global coordinate is
\be
ds^2=-d\tau^2+a^2(\tau)d^{d-1}\Omega_3^2\,,\quad a=\fft{\cosh (H_0 \tau)}{H_0}\,.
\ee
It is useful to define ``conformal time'' $\cosh(H_0 \tau)=-1/\sin(\eta H_0)$ \cite{Ivo:2024ill,Maldacena:2024uhs}, such that
\be
ds^2=\fft{1}{\sin^2(\eta H_0)}\left(-d\eta^2+\fft{d\Omega_3^2}{H_0^2}\right)\,.
\ee
In this case, the no-boundary condition that selects the Hartle–Hawking state is the regularity at $\tau = \pm i\pi/2$ for the ket and bra, respectively, applying not only to the mode functions but also to the geometry, with $a(\pm i\pi/2) = 0$ \cite{hartle1983wave,Lehners:2023yrj,Ivo:2024ill,Maldacena:2024uhs}.

\subsection{Modes and propagators}

We can solve the wave equation in this coordinate to find the modes for massless inflaton \cite{Ivo:2024ill,Maldacena:2024uhs}
\begin{align}
&\phi(\eta,\Omega_3)=\sum_{Jm_1m_2} \left(a_{J m_1m_2} u_J^-(\eta) Y_{Jm_1m_2}(\Omega)+a^\dagger_{J m_1m_2} u_J^+(\eta) Y^\ast_{J m_1m_2}(\Omega)\right)\,,\nonumber\\
& u_J^\pm(\eta)=\fft{H_0}{\sqrt{2 J(J+1)(J+2)}}\left((1+\fft{J}{2})e^{\pm i J H_0 \eta}-\fft{J}{2} e^{\pm i(J+2)H_0 \eta}\right)\,,\label{eq: global modes}
\end{align}
where $Y_{Jm_i}$ is the spherical harmonics satisfying
\be
\int d^{d-1}\Omega Y_{J m_i}Y_{J' m_j}^\ast=\delta_{J J'}\delta_{m_i m_j}\,,\quad \sum_{J m_i}Y_{Jm_i}(\Omega)Y_{J m_i}(\Omega')=\delta^{d-1}(\Omega-\Omega')\,.
\ee
We have the quantization condition:
\be
[\phi(\eta,\Omega),\Pi(\eta,\Omega')]=i\fft{\delta^{3}(\Omega-\Omega')}{H_0^3}\,,\quad [a_{Jm_i},a^\dagger_{J'm_j}]=\delta_{JJ'}\delta_{m_i m_j}\,.
\ee
Besides, we can construct the partial wave to decompose the Green functions
\begin{align}
& P_J^{d}(z)=\fft{\Omega_{d-1}}{{\rm dim}\rho_J^d}\sum_{m_i} Y_{J m_i}(\Omega_1) Y^\ast_{J m_i}(\Omega_2)=\, _2F_1\left(-J,d+J-2;\frac{d-1}{2};\frac{1-z}{2}\right)\,,\quad z=|\Omega_1-\Omega_2|=X_1\cdot X_2\,,\nonumber\\
& {\rm dim}\rho_J^d=\frac{(d+2 J-2) \Gamma (d+J-2)}{\Gamma (d-1) \Gamma (J+1)}\,,\quad P_J^d(x)=\fft{(-1)^J}{2^J \left(\fft{d-1}{2}\right)_J}(1-x^2)^{\fft{3-d}{2}}\fft{d^J}{dx^J}\left(1-x^2\right)^{J+\fft{d-3}{2}}\,,
\end{align}
where $|\Omega_{ij}|$ measures the distance on the sphere.
We take $d=4$ in formulas above. 

We can easily verify the validity of the flat-slice limit $J\rightarrow\infty, H \eta\rightarrow 0$ but fixed $J H \eta=k\eta$: taking this limit for the modes in \eqref{eq: global modes} reproduces the flat-slice solution in \eqref{eq: flat modes}.

We now decompose the correlators in the partial wave basis, which we denote as $G_J$
\be
\langle \phi(\eta_1,\Omega_1)\phi(\eta_2,\Omega_2)\rangle_{R/A/K}=\sum_{J} n_J^d G_{J}^{R/A/K}(\eta_1,\eta_2)P_J^d(z_{12})\,,
\ee
where
\be
n_J^d=\fft{{\rm dim}\rho_J^d}{\Omega_{d-1}}=\frac{2^{1-d} \pi ^{\frac{1}{2}-\frac{d}{2}} (d+2 J-2) \Gamma (d+J-2)}{\Gamma \left(\frac{d}{2}-\frac{1}{2}\right) \Gamma (J+1)}\,.
\ee
We thus find that the propagators are given by:
\begin{align}
& G_{J}^K(\eta_1,\eta_2)=H_0^2 \tfrac{(J+2)^2 \cos \left(\eta _{12} H_0 J\right)+J \left(J \cos \left(\eta _{12} H_0 (J+2)\right)-2 (J+2) \cos \left(\left(\eta _1+\eta _2\right) H_0\right) \cos \left(\eta _{12} H_0 (J+1)\right)\right)}{8 J (J+1) (J+2)}\,,\nonumber\\
& G_J^R(\eta_1.\eta_2)=-i H_0^2 \tfrac{(J+2)^2 \sin \left(\eta _{12} H_0 J\right)+J \left(J \sin \left(\eta _{12} H_0 (J+2)\right)-2 (J+2) \cos \left(\left(\eta _1+\eta _2\right) H_0\right) \sin \left(\eta _{12} H_0 (J+1)\right)\right)}{4 J (J+1) (J+2)}\,.
\end{align}
We can easily verify that their flat-slice limit precisely reproduces the propagators in the flat-slicing in \eqref{eq: propagators}, where we note
\be
n_J^d P_J^d(z)\rightarrow \fft{k^2 H_0^2}{2\pi^2} \fft{\sin (k |x|)}{k |x|}\,,
\ee
for $J=kH_0\rightarrow\infty$, $|\Omega_{12}|=|x| H_0 \rightarrow 0$ but $J |\Omega_{12}|= k |x|$ fixed.

\subsection{Coarse-graining and Fokker-Planck equation}

Let's now understand the coarse graining procedure in the global slicing. We consider tracing out the environment of modes with
\be
J>\varepsilon a H_0\gg 1\,,
\ee
after which we eventually consider $\varepsilon \ll 1$. Note that this condition for $J$ is not the flat-slice limit $\eta H_0\rightarrow 0$. Nevertheless, we note that taking $J>\epsilon a H_0 \gg1$ simultaneously with $\epsilon \ll 1$ is equivalent to the flat-slice limit. However, in our prescription, we first integrate out the short modes with large-$J$ limit, and then flow to $\varepsilon\rightarrow 0$, which retains the shape of the sphere without flattening it.

We now compute the leading diffusion coefficient. We find that our coarse-graining procedure gives
\begin{align}
&\langle\phi_e^c(\tau,\Omega)\phi_e^c(\tau',\Omega')\rangle= \sum_{J} \theta\left(J+\frac{\varepsilon}{\sin(\eta H_0)}\right)\theta\left(J+\frac{\varepsilon}{\sin(\eta' H_0)}\right) G^{cc}_J(\eta,\eta') P_j^d(z) n_J^d\nonumber\\
& \simeq \int dJ \fft{J^2}{2\pi^2} \theta\left(J+\frac{\varepsilon}{\sin(\eta H_0)}\right)\theta\left(J+\frac{\varepsilon}{\sin(\eta' H_0)}\right)  \lim_{J\rightarrow\infty}G_J^{cc}(\tau,\tau') \fft{\sin(J\sqrt{2(1-z)})}{J \sqrt{2(1-z)}}\,.\label{eq: global Gee}
\end{align}
Inserting the vertex, we can then use \eqref{eq: global Gee} to compute $\begin{tikzpicture}
[baseline={(0,-0.5ex)}]
\draw[thick,dashed] (-0.5,0.5)--(0,0);
\draw[thick] (0,0)--(1,0);
\draw[thick,->] (0,0)--(0.6,0);
\draw[thick,dashed] (1,0)--(1.5,0.5);
\node at (-0.05,0.25) {$q$};
\node at (1+0.05,0.25) {$q$};
\node at (0.15,0.15) {$c$};
\node at (0.85,0.15) {$c$};
\node at (0,-0.3) {$t_1$};
\node at (1,-0.3) {$t_2$};
\end{tikzpicture}$. From \eqref{eq: global Gee},
it is straightforward to see that the leading diffusion term essentially the same as in flat-slice, with the only difference coming from that the shape of the universe is a sphere
\be
C_{1}=\fft{H_0^3}{4\pi^2}{\rm tanh}\left(H_0 \tau\right)=\fft{H_0^2}{4\pi^2}\fft{\sqrt{H_0^2 a^2-1}}{a}\,.
\ee

We can now derive the FP equation. The Liouvelle Hamiltonian at leading order is
\be
H_L^{\rm eff}=\frac{\Pi^c \Pi^q}{a^3}+a^3 \phi^q v_r'(\phi^c)-iC_{1}(\Pi^q)^2\,.
\ee
To find the saddle-point, we set $\phi^q=0$ at time $t$ (then slow-roll limit of the equation for $\phi^q$ sets $\phi^q=0$ everywhere) and solve the equation of motion for $\phi^c$
\be
\ddot{\phi}^c+3H\dot{\phi}^c+v_r'(\phi^c)=0\,,
\ee
with $H=\dot{a}/a=H_0 {\rm tanh} (H_0\tau)$. At the leading slow-roll expansion $\ddot{\phi}^c=0$, we find
\be
\Pi^c_{\rm sad}=-\fft{a^4}{3\sqrt{a^2 H_0^2-1}} v_r'(\phi^c)\,.
\ee
This then leads to the FP equation \eqref{eq: FP global}.

Our method allows us to systematically implement this procedure to find higher-order diffusion terms and thus corrections to the FP equation and its phase space generalization. Nevertheless, we leave this for future studies.

\end{document}